\documentclass[preprint]{mn2e}

\usepackage{epsfig}

\newcommand{\Msun}{\mbox{\,M$_{\odot}$}}

\newcommand{\Rsun}{\mbox{\,R$_{\odot}$}}

\newcommand{\ergss}{\mbox{\,ergs\,s$^{-1}$}}
\newcommand{\Msyr}{\Msun\,\mbox{yr$^{-1}$}}
\newcommand{\yr}{\mbox{\,yr}}

\title[The Formation and Evolution of Black-Hole Binaries]
{On the formation and evolution of black-hole binaries}

\author[Podsiadlowski, Rappaport {\rm\&} Han]
{Ph.~Podsiadlowski$^1$\thanks{E-mail: podsi@astro.ox.ac.uk}, S.~Rappaport$^2$
\& Z.~Han$^3$\\
\it
$^1$ University of Oxford, Department of Astrophysics, Oxford,
OX1 3RH\\
$^2$ Department of Physics and Center for Space Research, Massachusetts
Institute of Technology, Cambridge, MA 02139, USA\\
$^3$ Yunnan Observatory, National Astronomical Observatories,
the Chinese Academy of Sciences, Kunming, 650011, China\\
}

\date{\today}

\volume{000}

\pagerange{\pageref{firstpage}--\pageref{lastpage}} \pubyear{2002}

\begin{document}

\maketitle

\label{firstpage}

\begin{abstract}

We present the results of a systematic study of the formation and
evolution of binaries containing black holes and normal-star
companions with a wide range of masses. We first reexamine the
standard formation scenario for close black-hole binaries, where the
progenitor system, a binary with at least one massive component,
experienced a common-envelope phase and where the spiral-in of the
companion in the envelope of the massive star caused the ejection of
the envelope. We estimate the formation rates for different companion
masses and different assumptions about the common-envelope structure
and other model parameters. We find that black-hole binaries with
intermediate- and high-mass secondaries can form for a wide range of
assumptions, while black-hole binaries with low-mass secondaries can
only form with apparently unrealistic assumptions (in agreement with
previous studies).

We then present detailed binary evolution sequences for black-hole
binaries with secondaries of 2 to 17\Msun\ and demonstrate that in
these systems the black hole can accrete appreciably even if accretion
is Eddington limited (up to 7\Msun\ for an initial black-hole mass of
10\Msun) and that the black holes can be spun up significantly in the
process.  We discuss the implications of these calculations for
well-studied black-hole binaries (in particular GRS 1915+105) and
ultra-luminous X-ray sources of which GRS 1915+105 appears to
represent a typical Galactic counterpart.  We also present a
detailed evolutionary model for Cygnus X-1, a massive black-hole
binary, which suggests that at present the system is most likely in a
wind mass-transfer phase {\em following} an earlier Roche-lobe
overflow phase. Finally, we discuss how some of the assumptions in the
standard model could be relaxed to allow the formation of low-mass,
short-period black-hole binaries which appear to be very abundant in
Nature.

\end{abstract}

\begin{keywords}
binaries: close -- stars: black holes -- X-rays: stars -- stars:
individual: GRS 1915+105 -- stars: individual: Cygnus X-1 -- gravitation
\end{keywords}

\section{Introduction}
There are currently 17 binary systems containing black holes for which
dynamical mass estimates are available (see e.g. Table 1 of Lee, Brown
\& Wijers 2002 [LBW], and references therein; Orosz et al.\ 2002).
According to conventional wisdom, these systems formed from primordial
binaries where at least one of the stars was quite massive (i.e. $M
\ga 20$\,--\,$25\Msun$).  If mass transfer from the primary to the
secondary commences at an orbital period in the range of $\sim$
1\,--\,10\yr, a common envelope may form during which the hydrogen-rich
envelope of the primary is expelled (Paczy\'nski 1976). If the
secondary and the core of the primary avoid a merger, then the massive
core may evolve to core collapse and the formation of a black hole in
a close binary.

For 9 of the 17 black-hole binaries (see e.g. LBW), the current-epoch
companion mass is $\la 1\Msun$ and the orbital periods are $\la 1$
day.  For reasons discussed later in the text, these systems probably
had primordial secondaries whose mass was not substantially greater
than $\sim 1.5 \Msun$ (but see \S~4.4).  One quantitative difficulty
with the common-envelope scenario for forming this type of black-hole
binary is that the amount of orbital energy that can be released by
the spiral-in of a low-mass secondary may not be sufficient to eject
the massive envelope of the primary.  It has long be recognized that
this is energetically challenging even if the common-envelope ejection
mechanism is very efficient (Podsiadlowski, Cannon \& Rees 1995;
Portegies Zwart, Verbunt \& Ergma 1997; Kalogera 1999; see, however,
also Romani 1992). Furthermore, recent determinations of the binding
energy of the envelopes of massive supergiants by Dewi \& Tauris
(2000, 2001) suggest that all studies so far may have significantly
underestimated how tightly bound these envelopes actually are, which
seriously aggravates the problem.  On the other hand, it has been
estimated that there may be up to several thousand low-mass black-hole
transients in the Galaxy (Wijers 1996; Romani 1998). This has led to
several alternative formation scenarios for low-mass black-hole
binaries, where either the low-mass companion is a third star in a
triple system being captured into a tight orbit when the two massive
components merge (Eggleton \& Verbunt 1986), or where the low-mass
star forms after the black hole -- out of a collapsed massive envelope
(Podsiadlowski et al.\ 1995). In the present work we reexamine the
standard formation scenarios for low-mass black hole binaries with
plausible modifications to some of the usual assumptions

By contrast, for 4 of the black-hole binaries, the mass of the
companion is substantially larger (i.e. $\ga 6 \Msun$), and the
availability of orbital binding energy for ejecting the common
envelope is greatly enhanced.  The remaining 4 systems (4U 1543-47, GRO
J1655-40, GS 2023+338, and GRS 1915+105) have either intermediate-mass
donor stars (i.e. $2 \la M_{\rm d} \la 5\Msun$) or orbital periods longer
than 2.5 days, thereby allowing for primordial secondaries of at least
intermediate mass, and substantial mass loss or evolution of the
secondary to its present status as the donor star. It is on the
evolution of these latter two categories, with particular emphasis on
GRS 1915+105, that we focus this work (for other recent discussions of
intermediate-mass black-hole binaries see Kalogera 1999; Brown et al.\
2000; LBW).

In addition to the common-envelope ejection mechanism, another
major uncertainty in the modelling of black-hole binaries is
the initial mass of the black hole which is caused by uncertainties in
the theory of both single and binary stellar evolution. Some of the
key factors that determine the maximum initial black-hole mass are (1)
the minimum initial mass above which a star leaves a black-hole
remnant (mostly believed to be in the range of 20\,--\,25\Msun; Maeder
1992; Woosley \& Weaver 1995; Portegies Zwart et al.\ 1997; Ergma \&
Fedorova 1998; Ergma \& van den Heuvel 1998; Brown et al.\ 2000; Fryer
\& Kalogera 2001; Nelemans \& van den Heuvel 2001; cf Romani 1992),
(2) the minimum
mass above which a single star loses its envelope in a stellar wind
and becomes a helium/Wolf-Rayet star, (3) the maximum radius of a
single star before and after helium core burning, (4) the amount of
wind mass loss in the Wolf-Rayet phase and (5) the fraction of the
mass that is ejected when the black hole forms (for detailed recent
discussions see Brown et al.\ 2000; Fryer \& Kalogera 2001; Nelemans
\& van den Heuvel 2001). Generally one expects the most massive black
holes to form from stars that have an initial mass close to the
minimum mass above which a star loses its hydrogen-rich envelope in a
stellar wind and becomes a Wolf-Rayet star, and where the
common-envelope phase occurs near the end of the evolution of the
massive primary (i.e. experiences case C mass transfer; Brown, Lee \&
Bethe 1999; Wellstein \& Langer 1999).  This avoids a long phase where
the mass of the helium star, the black-hole progenitor, is reduced by
a powerful stellar wind, as typically seen from Wolf-Rayet stars,
which would reduce the final helium-star mass and hence the maximum
black-hole mass (see e.g. Woosley, Langer \& Weaver 1995)\footnote{It
should also be noted that in the formation of some black holes
(e.g. the black hole in Nova Scorpii) a significant fraction of the
mass of the helium star is ejected in the supernova explosion in which
the black hole formed (Podsiadlowski et al.\ 2002; LBW). Thus the final
helium-star mass strictly provides only an upper limit on the
black-hole mass.}. Unfortunately, the evolutionary tracks for massive
post-main-sequence stars and in particular the maximum radius a
star attains after helium core burning are rather uncertain (and
generally inconsistent with observed distributions of stars in the
Hertzsprung-Russell diagram; see e.g. Langer \& Maeder 1995).  Fryer
\& Kalogera (2001) have shown that initial black-hole masses as high
as 15\Msun\ can be obtained if either the parameter range for case C
mass transfer is increased or the wind mass-loss rate in the
helium-star phase of the black-hole progenitor is reduced
(see also Brown et al.\ 2001; Nelemans \& van den Heuvel 2001;
Belczynski \& Bulik 2002; Pols \& Dewi 2002)

Only a few of the previous studies (e.g. LBW) have considered the
possibility that the black hole may increase its mass substantially
since its formation by mass transfer from the companion star. It is
one of the purposes of this paper to demonstrate that accretion from a
companion star can substantially increase the mass of a black hole and
spin it up in the process and that the present mass may not be
representative of the initial black-hole mass.  A closely coupled
result is that the observed donor star masses may be substantially
lower than their initial mass.

The paper is structured as follows. In \S~2 we present detailed binary
population synthesis calculations to show how the formation rate of
black-hole binaries and the distribution of the secondary masses
depend on the structure of massive supergiant envelopes and the
modelling of common-envelope ejection. In \S~3 we discuss the results
of extensive binary evolution calculations for black-hole binaries with
intermediate-/high-mass secondaries, which we then apply in \S~4 to
observed systems, in particular GRS 1915+105, ultraluminous X-ray
sources and Cyg X-1.  Finally, in \S~4 we reexamine the standard
formation scenario for low-mass black-hole binaries to
understand why such systems appear to be so plentiful in Nature.

\section{Binary Population Synthesis Model}

\noindent
\subsection{Assumptions of the Model}

In this work we consider only black-hole binaries that descend from a
primordial binary pair of stars.  Black-hole binaries that may form
dynamically in globular clusters are left for another study.  We start
our investigation of black-hole binaries, which includes (initially)
intermediate- and high-mass donors, by carrying out a Monte Carlo
population study aimed at producing systems at the end of the
common-envelope phase.  This will provide guidance for the second part
of our study where we follow in detail the X-ray binary phase which
involves mass transfer from the donor star to the black hole.

The assumptions and ingredients that we adopt for the population study are
listed below.  For prior studies of the formation and evolution of black
hole binaries see e.g. Romani (1992), Portegies Zwart, Verbunt \& Ergma
(1997), Ergma \& van den Heuvel (1998), Ergma \& Fedorova (1998), Kalogera
(1999), Brown et al.\ (2000) and Fryer \& Kalogera (2001); also see
Kalogera \& Webbink (1998) for a related, detailed study of the theoretical 
constraints on the formation of neutron-star X-ray binaries.

We utilize a simple power-law distribution for the initial mass
function (IMF) for the primary stars in primordial binaries.
Specifically, we take $dN/dM_{\rm p} \propto M_{\rm p}^{-x}$ with $x = 2.35$
(Salpeter 1955).  Since we consider a relatively small range of
primary masses, the results are not very sensitive to the particular
choice of $x$ or to the fact that the IMF flattens significantly
toward lower masses (see e.g. Miller \& Scalo 1979). Only primordial
binaries with mass $M_{\rm p} > 25\Msun$ are considered as progenitors of
black holes.  This lower limit is somewhat uncertain, but is
consistent with current models of massive stars and the modelling of
supernova explosions (e.g.  Woosley \& Weaver 1995; Fryer \& Kalogera
2001).  We also somewhat arbitrarily take the upper mass limit for the
primary to be $M_{\rm p} < 45~\Msun$.

The mass of the secondary star, $M_{\rm s}$, in the primordial binary
is chosen from a flat mass ratio distribution, i.e. $f(q)=1$, where
$q \equiv M_{\rm s}/M_{\rm p}$. There is a large degree of uncertainty
in the actual distribution of mass ratios in binaries; however, our
choice reflects the simple fact that a significant fraction of high-mass
stars are observed to have high-mass companions (see e.g.
Garmany, Conti \& Massey 1980). At the low-mass end, only secondaries
with mass $M_{\rm s} \ga 0.5\Msun$ are retained in the population synthesis.

The orbital-period distribution of primordial binaries is taken to be
constant in $\log P_{\rm orb}$, where $P_{\rm orb}$ is the orbital period
(see e.g. Abt \& Levy 1978).  While orbital eccentricity among primordial
binaries might be expected to have a distribution that increases linearly
with $e$ (Duquennoy \& Mayor 1991), we have simply taken all orbits to be
circular.  One motivation for this choice is that the tidally circularized
orbital radius of an eccentric binary is $(1-e^2)a$, which we note is
simply linearly proportional to $a$, the initial orbital semimajor axis.
Thus, the distribution in circularized orbital radii would be the same as
that for the semimajor axes, regardless of the distribution in $e$, as long
as that distribution is independent of $a$.

The evolution of the primary star as it expands toward filling its Roche
lobe is followed with the Hurley, Pols \& Tout (2000, hereafter HPT) code.
The evolution includes wind mass loss from the primary according to the
prescription of Nieuwenhuijzen \& de Jager (1990).  These stars have fully
developed core masses given by $M_{\rm core} \simeq 0.12~M_{\rm p}^{1.35}$
(in solar masses; see e.g. HPT).
Typical wind mass losses prior to the primary traversing
the Hertzsprung gap (hereafter HG) amount to only $\sim 2-6\,\Msun$;
however, much larger losses can be sustained once stellar radii of $\ga
1000\Rsun$ are reached.  In this latter regard, we compute the orbital
widening due to the stellar wind mass loss according to $da/a =
-dM_{\rm w}/M_{\rm b}$, where the fractional change in orbital
separation is equal to the wind mass loss from the primary ($dM_{\rm w}$)
divided by the total mass of the binary, $M_{\rm b}$.  This expression is
based only on the assumption that the specific angular momentum carried
away by the wind is equal to that of the primary star in the binary orbit
(see, however, the discussion in \S~4.4).

An interesting possibility occurs if the initial Roche-lobe radius,
$R_{\rm L}$, of the primary falls in the range $R_{\rm HG} < R_{\rm L}
< R_{\rm max}$, where $R_{\rm HG}$ is the stellar radius at the end of
the Hertzsprung gap, and $R_{\rm max}$ is the maximum radius that the
primary can attain.  It is in this phase of the evolution that the
primary can lose a substantial fraction of its envelope mass in a
stellar wind, thereby requiring less orbital binding energy to eject
the remaining envelope of the primary once mass transfer commences
(see the discussion below).  However, as the primary expands and loses
mass in a wind, the mass loss also causes the orbit to expand (as
described above).  It it therefore not obvious whether the expanding
star can catch up with its Roche lobe.  To quantify this issue, we
first define a Roche-lobe index due to wind loss:
\begin{equation}
\xi_{\rm L,w} \equiv \left(\frac{d\ln R}{d\ln M_{\rm b}}\right)_{\rm L,w} \simeq
1-\left(\frac{0.087}{r_{\rm L}}\right)\left(\frac{M_{\rm b}}{M_{\rm p}}\right)
\end{equation}
as well as a stellar index associated with wind mass loss:
\begin{equation}
\xi_{\rm *,w} \equiv \left(\frac{d\ln R}{d\ln M_{\rm b}}\right)_{\rm *,w} \simeq
\left(0.023 + 0.00086\left[30-M_{\rm HG}\right]\right)M_{\rm b}
\end{equation}
where $M_{\rm b}$ is the instantaneous total mass of the binary
(including wind mass loss), $M_{\rm HG}$ is the mass of the primary at
the end of the Hertzsprung gap (both in solar masses), and $r_{\rm L}$
is the Roche-lobe radius of the primary in units of the orbital
separation. In both cases, a minus sign has been subsumed into the
definition of $dM_w$ since it is always negative.  The above
expressions for $\xi_{\rm *,w}$ and $\xi_{\rm L,w}$ were derived from
fits to the stellar models in HPT and the expression for the
Roche-lobe radius of Plavec (1968), respectively. If $\xi_{\rm *,w} <
\xi_{\rm L,w}$ when the primary expands past the HG, then it will
never catch up with its Roche lobe since $\xi_{\rm *,w}$ decreases
faster with mass loss than does $\xi_{\rm L,w}$ (this catch-up problem
has first been identified by Kalogera \& Webbink [1998] in their study
of the formation of neutron-star X-ray binaries).  We have found that
even if the reverse inequality holds when the primary expands past the
HG, it is still extremely rare for the star to overtake its Roche
lobe.

If the primary does evolve to the point of overflowing its Roche lobe
(almost always before the end of the HG), we use a simple prescription for
deciding whether a common-envelope phase ensues.  If the primary has
evolved at least to the start of the Hertzsprung gap and the mass ratio
$M_{\rm p}/M_{\rm s}$ exceeds 2.0, or if the primary is beyond the
HG and the mass
ratio exceeds a value of 1.2, we assume that a common envelope will occur.
For other conditions we take the mass transfer to be quasi-conservative and
stable.  These latter systems are not very common and do not lead to the
type of black-hole binary that we are considering.

\begin{figure*}
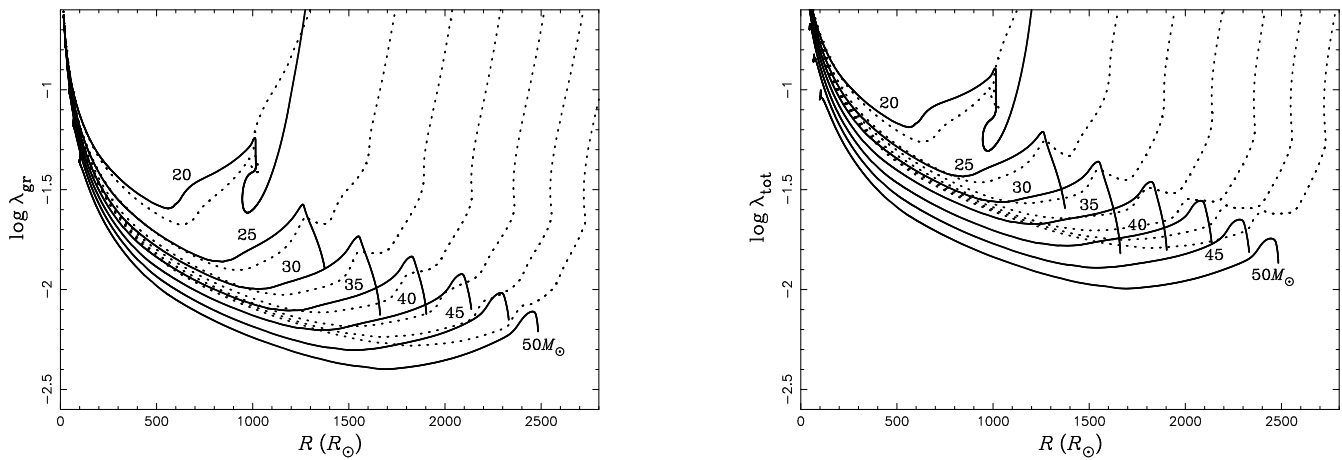

\psfig{file=Fig1a.ps,width=6cm,angle=-90}\hfill
\psfig{file=Fig1b.ps,width=6cm,angle=-90}
\caption{The envelope structure parameter $\lambda$ as a function
of stellar radius for different masses as indicated after hydrogen has
been exhausted in the core. In the panels on the left, $\lambda$ only
includes the gravitational binding energy, while on the right
$\lambda$ includes both the gravitational and the thermal energy
(similar to Dewi \& Tauris 2000). The dotted curves are calculated
without inclusion of a stellar wind. Note that in this case $\lambda$
always becomes large at the largest radii attained by the models.
}\label{Fig1}
\end{figure*}

A commonly used prescription was employed
to determine the orbital separation after
the common-envelope phase (see e.g. de Kool 1990; Dewi \& Tauris 2000):
\begin{equation}
\left( \frac{a_{\rm f}}{a_{\rm i}} \right)_{\rm CE} = \frac{M_{\rm c}M_{\rm s}}{M_{\rm p}} \left(M_{\rm s} +
\frac{2M_{\rm e}}{\alpha_{\rm CE}\lambda r_{\rm L}} \right)^{-1}
\end{equation}
where the subscripts $f$ and $i$ denote the final and initial values,
respectively, $M_{\rm c}$ and $M_{\rm e}$ are the core and envelope masses of the
primary, respectively, $\alpha_{\rm CE}$ is the efficiency with which the
orbital binding energy can liberate the envelope of the primary,
$\lambda^{-1}$ is the energy of the envelope of the primary
in units of $-GM_{\rm p}M_{\rm e}/R_{\rm p}$, 
and $r_{\rm L}$ is the dimensionless Roche-lobe radius of the primary.

Appropriate values of the $\lambda$ parameter are derived from stellar
structure and evolution calculations (with similar assumptions as in
HPT and wind mass loss according to Nieuwenhuijzen \& de Jager 1990)
that we have carried out specifically for this work.  These results
are shown in Figure~\ref{Fig1}, where the definition for $\lambda$ in the
left panel includes only the gravitational binding energy ($\lambda_{\rm
gr}$), while in the panel on the right it also includes the thermal
and the ionization energy ($\lambda_{\rm tot}$; see Han, Podsiadlowski
\& Eggleton 1994; Dewi \& Tauris 2000). We have compared these curves
with those computed by Dewi \& Tauris (2000, 2001) and found them to
be in reasonable agreement at similar evolutionary stages, although
the evolutionary tracks are slightly different. We note that the value
of $\lambda$ itself depends to some degree on the definition of the
core -- envelope boundary (see Han et al.\ 1994 and Tauris \& Dewi
2001). Here we have defined the core mass as the central mass that
contains 1\Msun\ of hydrogen (in contrast Dewi \& Tauris [2000]
defined it as the central mass which includes 10 per cent of the total
mass of hydrogen). In agreement with Tauris \& Dewi (2001) we find
that in some evolutionary phases the value of $\lambda$ can be a
rather sensitive function of the chosen core -- envelope boundary;
this is particularly true for very evolved supergiants near the end of
their evolution. Unfortunately, this uncertainty in the definition of
$\lambda$ cannot easily be resolved without a better understanding of
the CE ejection process.

Values of $\lambda_{\rm tot}$ that are appropriate to stars in the
mass range $25-45\Msun$ and the later phases of their evolution
(i.e. in or beyond the HG) lie in the range of $0.01 \la
\lambda \la 0.06$.  Note that for a 20\Msun\ model $\lambda$ increases
to a value $\sim 1$ near the very end of the evolution. This occurs
when the star ascends the asymptotic-giant branch and develops a deep
convective envelope, creating a steep chemical gradient
($\mu$-gradient) below the convective envelope (i.e.\ establishes a
typical giant structure). The more massive stars experience a
supernova before this phase is reached. Since these results depend on
the assumptions in the stellar modelling, we also performed a series
of calculations without any wind mass loss, shown as dotted curves in
Figure~\ref{Fig1}. The models now achieve significantly larger radii,
and $\lambda$ increases dramatically near the end of the evolution in
all cases. This demonstrates that the behaviour of $\lambda$ is also
quite sensitive to uncertainties in the stellar modelling itself,
adding another uncertainty to the problem.  In view of these
uncertainties we adopted a value of $\lambda$ in this work that is
held constant throughout the evolution of a star. We were careful,
however, to test the full range of plausible values for $\lambda$
($0.01-0.5$).

The value for $\alpha_{\rm CE}$ was simply taken to be unity. Since
the two parameters $\alpha_{\rm CE}$ and $\lambda$ appear as a
product in the expression for the post-CE orbital separation, the
uncertainty in one can, to some degree, be incorporated into the
uncertainty in the other. A value of $\alpha_{\rm CE}\sim 1$ was
motivated by a number of recent empirical studies of the efficiency of
the CE ejection process which have suggested that the CE ejection
process must be very efficient; see e.g. the recent study of sdB stars
in compact binaries by Han et al.\ (2002a,b) (these are short-period
binaries which have formed through a very well defined CE channel and
are hence particularly suitable for studying the CE ejection process
empirically). In their best model, these authors found $\alpha_{\rm
CE} = 0.75$ and that $\sim 75$\, per cent of the thermal energy of the
envelope had to be used in the ejection process to explain the
observed orbital-period distribution of compact sdB binaries.
However, this empirical study, as well as all other previous ones,
strictly apply only to systems where the donor is a giant with a
convective envelope but not to systems which experience a CE phase
when the donor star is in the Hertzsprung gap and has a radiative
envelope. Such stars have fundamentally different internal structures
and are much more centrally concentrated than more evolved
(super-)giants with convective envelopes (see e.g. Fig.~2 in
Podsiadlowski 2001) (this is e.g. reflected in the low value of
$\lambda$ for stars in the Hertzsprung gap in Fig.~1). 
In such systems, the core -- envelope separation may not be distinct enough to
allow envelope ejection even if enough energy is available in
principle (see e.g. Taam \& Sandquist [2000] and Kalogera [2002; private
communication]); hence a value of $\alpha_{\rm CE}\sim 1$ may not be an
appropriate one for systems that experience a CE phase 
in the Hertzsprung gap.

Immediately after the common envelope phase has occurred, we check whether
the secondary star is overfilling its Roche lobe.  If so, we assume that
the secondary merges with the core and do not follow the binary further.

\begin{figure*}
\psfig{file=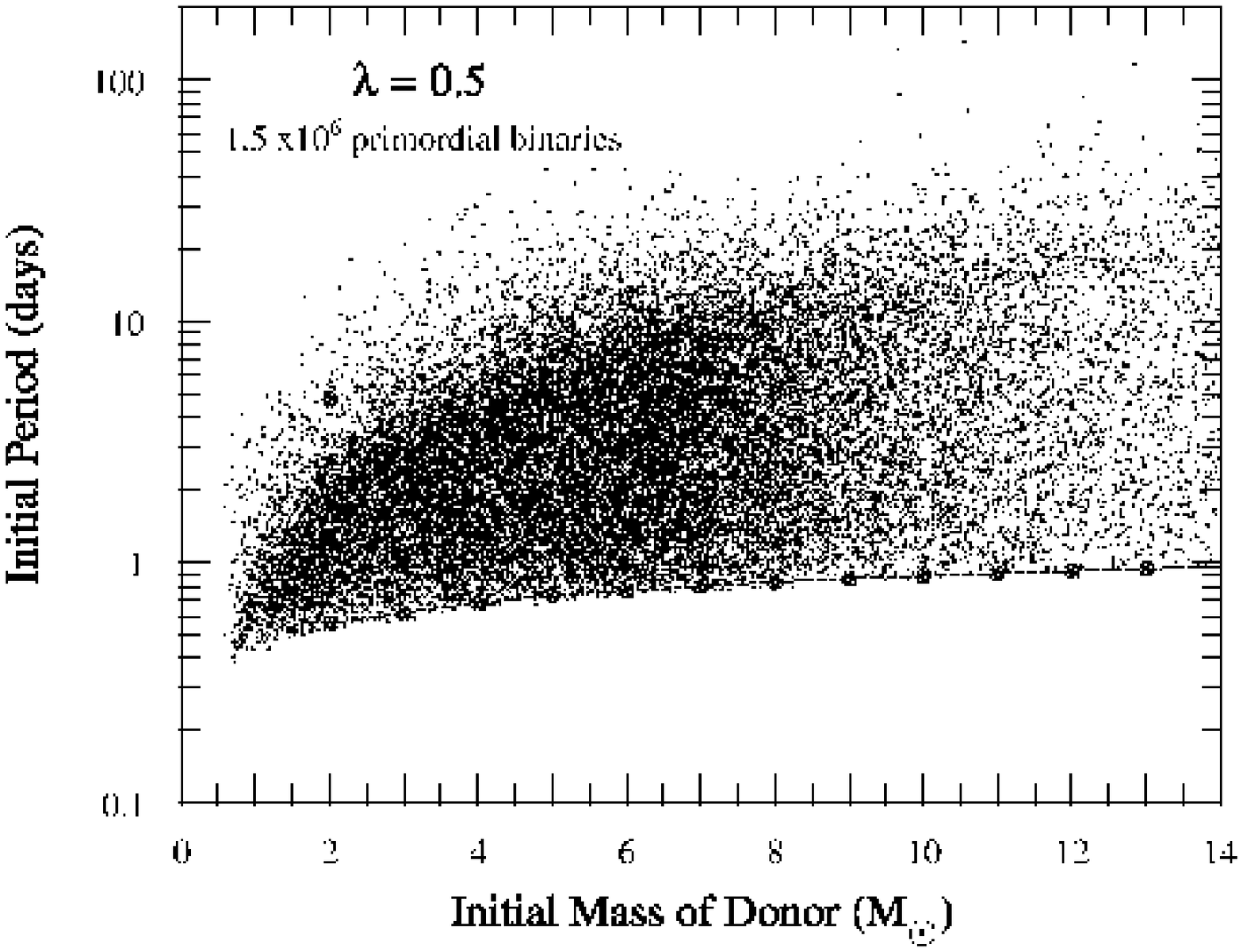,width=6cm,angle=90}\hfill
\psfig{file=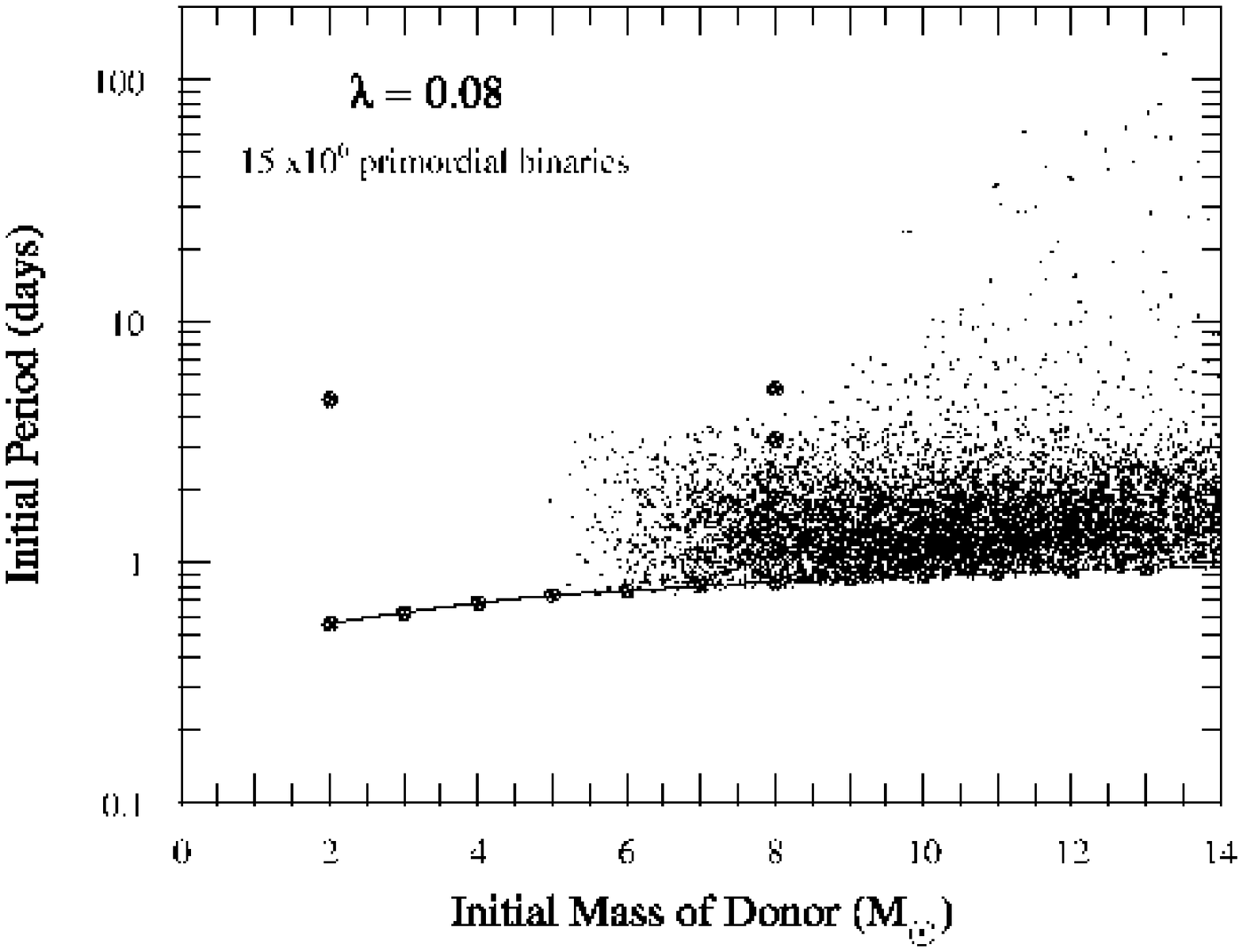,width=6cm,angle=90}
\caption{
Distribution of initial orbital period and initial secondary mass for
black-hole binaries at the beginning of mass transfer for different
assumptions about the common-envelope (CE) structure and ejection
efficiency, characterized by the parameters $\lambda$
and $\alpha_{\rm CE}$, respectively, where $E_{\rm env}^{\rm CE} =
GM_{\rm e}\,M_{\rm p}/\lambda\,R_{\rm p}$ is the
binding energy of the envelope.  The value of $\alpha_{\rm CE}$
was fixed as 1, while two different illustrative values of $\lambda$ were
used to produce the right and left panels. The solid circles indicate the
initial positions of our detailed binary evolution sequences which are
discussed in \S 3.}\label{Fig2}
\end{figure*}

The amount of wind mass loss from the exposed H-exhausted core of the primary
as it evolves toward core collapse is quite uncertain.  To cover a
full range of possibilities we adopt the following somewhat {\em ad
hoc} prescription for the mass of the black hole that eventually forms
from the collapse: $M_{\rm BH} = 6\Msun + (M_{\rm
He}-6\,\Msun)\mathcal{R}$, where $M_{\rm He}$ is the mass of the newly
exposed H-exhausted core, and $\mathcal{R}$ is a uniform random variable
between 0 and 1.  With this prescription we produce black holes of
mass over the range of $\sim 6 - 20\,\Msun$.  The orbital separation
at the end of this wind loss phase is taken to be a factor of $(M_{\rm
He}+M_{\rm s})/(M_{\rm BH}+M_{\rm s})$ larger than the separation at
the end of the common-envelope phase (see the expression above for the
change in orbital separation with wind mass loss). We specifically
chose this procedure to obtain black-hole masses that are reasonably
consistent with the observed ones (e.g. Fryer \& Kalogera 2001; van
den Heuvel 2001; LBW), but also to avoid some of the theoretical
uncertainties that lead to these masses (e.g. the wind mass-loss rate
in the helium-star phase, the question of case B vs.\ case C mass
transfer; see the discussion in \S~1). If it is indeed necessary that
systems with relatively massive black holes have experienced case C
mass transfer in the past (e.g.  Brown et al.\ 1999; Wellstein \&
Langer 1999; Langer 2002 [private communication]), our procedure
implicitly assumes that those evolutionary tracks for single stars
which presently do not allow case C mass transfer for stars more
massive than $\sim 20\Msun$ are not correct (also see Fryer \&
Kalogera 2001). Note also that our procedure does not produce
relatively low-mass black holes with masses $< 6\Msun$ which could have
a somewhat different evolution from the black-hole systems studied in
this paper (see Fryer \& Kalogera 2001 and Beer \& Podsiadlowski 2002).

When the evolving He star undergoes core collapse, we assume that the
entire mass of the He star collapses into the newly formed black hole
(but see LBW), and that there is no natal ``kick'' imparted to the
black hole.  Much has been inferred about natal kicks to neutron stars
from the proper motions of radio pulsars and from studies of numerous
individual binary systems containing neutron stars (see e.g.,
Lyne \& Lorimer 1994; Brandt \& Podsiadlowski 1995; Verbunt \& van
den Heuvel 1995; Hansen \& Phinney 1997; van den Heuvel et al.\ 2000;
Arzoumanian, Chernoff \& Cordes 2002).  Yet, relatively little is
understood about the physical origin of these kicks or the conditions
under which kicks are developed (see e.g. Janka \& M\"uller 1994; Fryer,
Burrows \& Benz 1997; Spruit \& Phinney 1998; Fryer \& Heger 2000;
Lai 2000; Pfahl et al.\ 2002).  Unfortunately, even less is
known about the nature of natal kicks delivered to forming black
holes.  The spatial distribution and the kinematics of the majority of
black-hole binaries appear to be consistent with the assumption that
no asymmetric kicks are imparted to the black hole (Brandt, Podsiadlowski
\& Sigurdsson 1995; White \& van Paradijs 1996), although there is
at least one clear exception (the black hole in Nova Sco; Brandt et al.\
1995; Fryer \& Kalogera 2001; Podsiadlowski et al.\ 2002a; for a
different view see Nelemans, Tauris \& van den Heuvel 1999).

In order to compute an absolute formation rate for black-hole binaries in
the binary population synthesis (hereafter BPS; see next section), we take
the birth rate of stars with mass $\ga 8\Msun$ to be 1 per 100
years, to match the Galactic supernova rate (of non-type Ia SNe; see e.g.
Cappellaro, Evans \& Turatto 1999).  This rate, coupled with the assumed
slope of the IMF, yields a maximum potential formation rate for black hole
primordial binaries of $\sim 6 \times 10^{-4}$ yr$^{-1}$; this includes an
approximate fraction for stars born in binaries of $\sim 1/2$.

Finally, we note that once the incipient black-hole binaries have been
produced, the luminous X-ray phase will not start until the secondary
has evolved to either produce a strong stellar wind or overflow its
Roche lobe.  In the BPS portion of this work we carry the calculations
only to the point of the successful ejection of the common envelope or
merger of the secondary with the core of the primary. We also apply a
very simplistic check for the dynamical stability of mass transfer
when the donor star ultimately commences mass transfer via Roche-lobe
overflow onto the black hole.  Here we required that $M_{\rm s}
\la M_{\rm BH}$; however, we note that in the detailed binary evolution
calculations presented in \S 3, the stability of the mass transfer is
calculated self-consistently in each evolution step, and that even
higher mass donor stars may result in stable mass transfer.  The
successful incipient black-hole binaries are tabulated and statistical
results are extracted.

Many of the above assumptions regarding the formation of black-hole
binaries are discussed in further detail by Romani (1992), Portegies Zwart,
Verbunt \& Ergma (1997), Ergma \& van den Heuvel (1998), Ergma \& Fedorova
(1998), Kalogera (1999), Brown et al.\ (2000), and Fryer \& Kalogera
(2001). One substantive difference in the assumptions that we make as
compared to those utilized in prior work is that we do {\em not} set
any {\em a priori} restrictive upper limit on the mass of the donor
star in the incipient black-hole binary.  Also, the values for
$\lambda$ that we use are considerably smaller (by up to a factor
of $\sim 20$) than the values employed in most earlier work.  Smaller
values of $\lambda$ require the expenditure of more orbital binding energy
to successfully eject the common envelope.  Finally, we do {\em not}
require that the donor star be unevolved when it commences mass transfer
onto the black hole (cf. Kalogera 1999).  This is due to the fact that we
explicitly include wide binaries containing black holes.

\subsection{Binary Population Synthesis Results}

\begin{figure*}
\centerline{\psfig{file=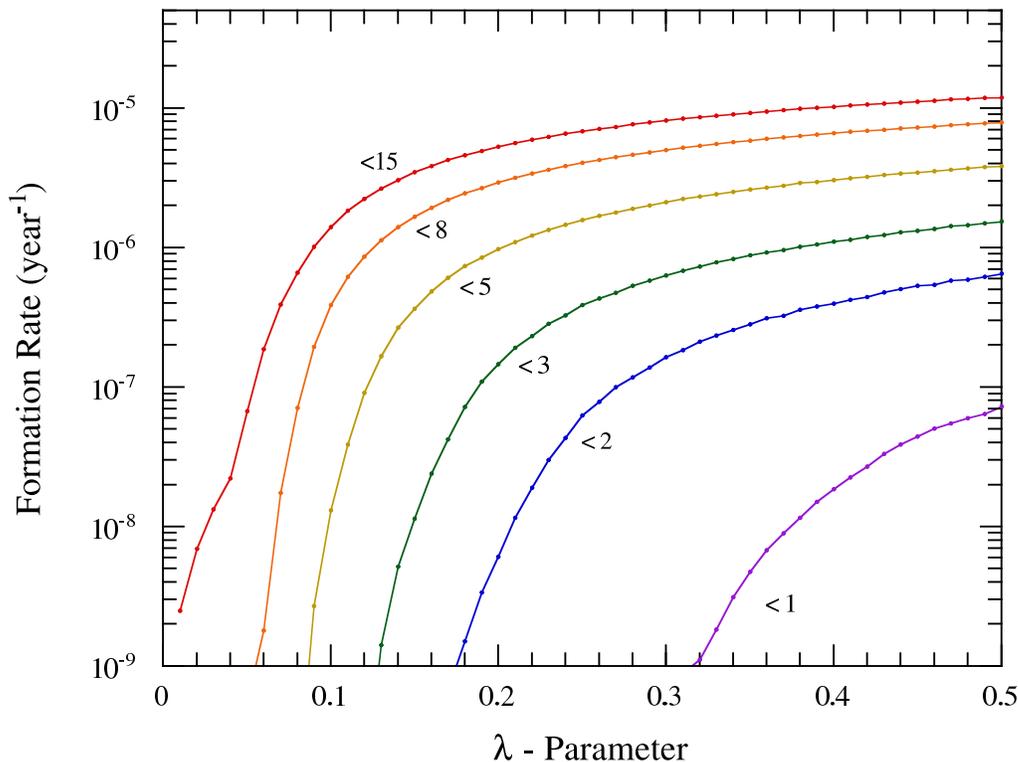,width=10cm,angle=90}}
\caption{Formation rate of black-hole binaries as a function
of the envelope structure parameter $\lambda$. The different
curves assume a different maximum mass for the secondary (in $\Msun$),
as indicated along the respective curves.}\label{Fig3}
\end{figure*}

Each binary population synthesis run (BPS) was typically started with $\sim
10^7$ primordial binaries, all of which have $M_{\rm p} > 25\Msun$.  Each
binary is followed using the prescriptions, assumptions, and algorithms
specified in the previous section, until it either becomes an incipient
black-hole X-ray  binary or goes down an alternate evolutionary path. The
relevant binary parameters are stored for each ``successful'' system.

Illustrative results from a BPS run with the $\lambda$-parameter set
equal to 0.5, a conventionally used value, are shown in
Figure~\ref{Fig2}a.  Each dot in the $P_{\rm orb}-M_{\rm d,i}$ plane
represents a single incipient black-hole X-ray binary just after the
black hole has been formed; $P_{\rm orb}$ is the orbital period and
$M_{\rm d,i}$ is the initial mass of the donor star.  The number of
primordial binaries used in this run was limited to only $1.5 \times
10^6$ in order to keep the density of points legible in the figure.
The sharp lower boundary in Figure~\ref{Fig2}a results from the fact that
systems with shorter orbital periods have merged, i.e. the donor star
overflowed its critical potential lobe at the end of the common-envelope
phase.  The filled circles are starting models for
the detailed evolutionary calculations of the binary X-ray phase
presented in \S~3.

In all, there are about 25,000 dots in Figure~\ref{Fig2}a, each
representing the successful formation of an incipient black-hole
binary X-ray source.  The expression we use to convert this number,
$N$, to a formation rate is as follows:
\begin{equation}
R = \left(\frac{1}{2}\right) \left(\frac{1}{8}\right)
\left(\frac{N}{N_0}\right) \tau_{\rm SN}^{-1}
\end{equation}
where $N_0$ is the starting number of primordial binaries (with $M_{\rm p} >
25\Msun$), the factor of 1/2 is an approximation to the fraction of
stars born in binaries, the factor of 1/8 represents the fraction of
stars capable of collapsing to a neutron star or black hole that comes
from the mass range $25-45\Msun$, and $\tau_{\rm SN}^{-1}$ is the Galactic
supernova rate (of non-type Ia SNe; see discussion above). The factor
of 1/8 depends somewhat, but not sensitively, on the slope of the IMF.
The particular example shown in Figure~\ref{Fig2}a then implies a
formation rate for such systems of $\sim 1 \times 10^{-5}$
yr$^{-1}$.

A second illustrative example of a BPS run is shown in
Figure~\ref{Fig2}b for the case where the $\lambda$-parameter has
been reduced to 0.08.  The number of primordial binaries in this
run was 10 times that used in Figure~\ref{Fig2}a.  In spite of this,
the number of systems represented in Figure~\ref{Fig2}b is actually
fewer than in Figure~\ref{Fig2}a due to the highly reduced formation
efficiency associated with the lower value of $\lambda$.
Note that, in addition
to the overall drop in formation efficiency, there is a complete
dearth of systems with donor masses of $\la 5\Msun$.  This results
from the fact that for small values of $\lambda$ (i.e. more envelope
binding energy) the systems with lower-mass donor stars do not have
sufficient orbital energy to unbind the common envelope.

We have systematically investigated the effect of $\lambda$ on the
formation efficiency of incipient black-hole binaries.  The BPS code
was run for 50 values of $\lambda$ in the range of $0.01-0.5$, in
steps of 0.01, and a formation efficiency computed for each.  The
results, shown in Figure~\ref{Fig3}, are further broken down according
to the donor masses contributing to the population.  The top curve
shows the formation rate for systems with all donor masses ($<
15\Msun$) computed according to equation (4).  Note the abrupt drop in
formation rate for values of $\lambda \la 0.1$.  The sequence of
curves below this is for the formation rate of black-hole X-ray
binaries with donor masses $<8, <5, <3, <2, {\rm and} <1\Msun$,
respectively.  The small shoulder on the curve for $<15~M_\odot$ at
very small values of $\lambda$ results from Roche-lobe overflow when
the primary has evolved beyond the HG and has lost a significant
amount of its envelope to a stellar wind.  Two immediate conclusions
to be drawn from the results in Figure~\ref{Fig3} are: (1) black-hole
binaries with low {\em incipient} donor masses (i.e. $\la 1\Msun$)
have extremely low formation rates unless either seemingly unphysical
(large) values of $\lambda$ are invoked or much lower-mass primary
stars are able to form black holes (see e.g. Portegies Zwart et
al.\ 1997; Kalogera 1999); and (2) the formation rates for black-hole
binaries with {\em incipient} intermediate-mass donor stars can be
quite substantial provided that the $\lambda$-parameter is not smaller
than $\sim 0.1$.

In addition to the BPS results shown in Figure~\ref{Fig3} we have
carried out some additional studies of how the theoretical
uncertainties in the values of $\lambda$ affect our conclusions.  As
can be seen from Figure~\ref{Fig1}, the values of $\lambda$ depend
systematically on the stellar mass, with the larger values (smaller
envelope binding energies) generally being associated with the lower
masses.  Since, to this point, we have excluded primary masses below
$25\Msun$, and these have the largest values of $\lambda$, we have
carried out an additional sequence of BPS calculations with the
inclusion of 20\,--\,25\Msun\ primaries to test whether these would
significantly enhance the formation rate of black-hole binaries.
Here, we simply recomputed the formation rates, as in
Figure~\ref{Fig3}, but now with the minimum primary mass set to
20\Msun.  Again, for the purposes of the BPS calculations, we took the
value of $\lambda$ to be fixed for primaries of all masses and
evolutionary states, and studied how the formation rates changed with
a systematic variation in the adopted value of $\lambda$.  The result
is that for any given value of $\lambda$, the rates are larger than
those shown in Figure~\ref{Fig3} by a factor typically limited to
$\sim 1.5$. This is a bit less than might be expected simply from the
increased primary mass range -- weighted by the Salpeter mass
function.  The smaller than anticipated increase is due to the fact
that many of the systems produced by $\sim 20\Msun$ primaries have
relatively low-mass black holes and the subsequent mass transfer is
unstable.

Finally, we constructed from the results of Figure~\ref{Fig1} a simple
dependence of the $\lambda$ parameter on primary mass: $\lambda =
\lambda_0 \exp (-0.074[M_{\rm p}-20\Msun])$, where $\lambda_0$ is a constant
to be supplied.  For the `plateau' regions in Figure~\ref{Fig1}b, the value
of $\lambda_0$ would be approximately 0.08.  We then produced a sequence
of BPS runs, including primary masses down to $20\Msun$, where the
value of $\lambda_0$ was varied systematically over the range of
0.01\,--\,0.5.  Thus, for any given choice of $\lambda_0$, in a given BPS
run, the values of $\lambda$ used scale with primary mass as in the
exponential expression given above.  The net result of this study, as
expected, is that the formation rates are all systematically {\em
reduced} compared to the values shown in Figure~\ref{Fig3} where a
fixed value of $\lambda$ was utilized in each BPS run.  The reason for
this reduction in formation rates is that now $\lambda_0$ represents
an upper limit to the value of $\lambda$ (corresponding to $20
\Msun$ stars), with all higher-mass stars having smaller values of
$\lambda$.

The conclusion drawn from these experiments, that extend the primary
masses down to 20\Msun\ and compute $\lambda$ from a prescription
that depends on the primary mass, is that the rates shown in
Figure~\ref{Fig3} are, to within a factor of a few, likely to be
{\em upper limits.}

\section{Binary Evolution Calculations}

In this section we present a series of binary stellar evolution
calculations to illustrate the evolution of black-hole binaries with
intermediate- and high-mass secondaries. In \S~3.1 we briefly describe
the binary evolution code and the main assumptions used in the
calculations, and in \S~3.2 we present the main results of the
calculations. In the subsequent section we will apply these results to
a variety of systems.

\subsection{Description of the code and principal assumptions}

\begin{figure*}
\centerline{\psfig{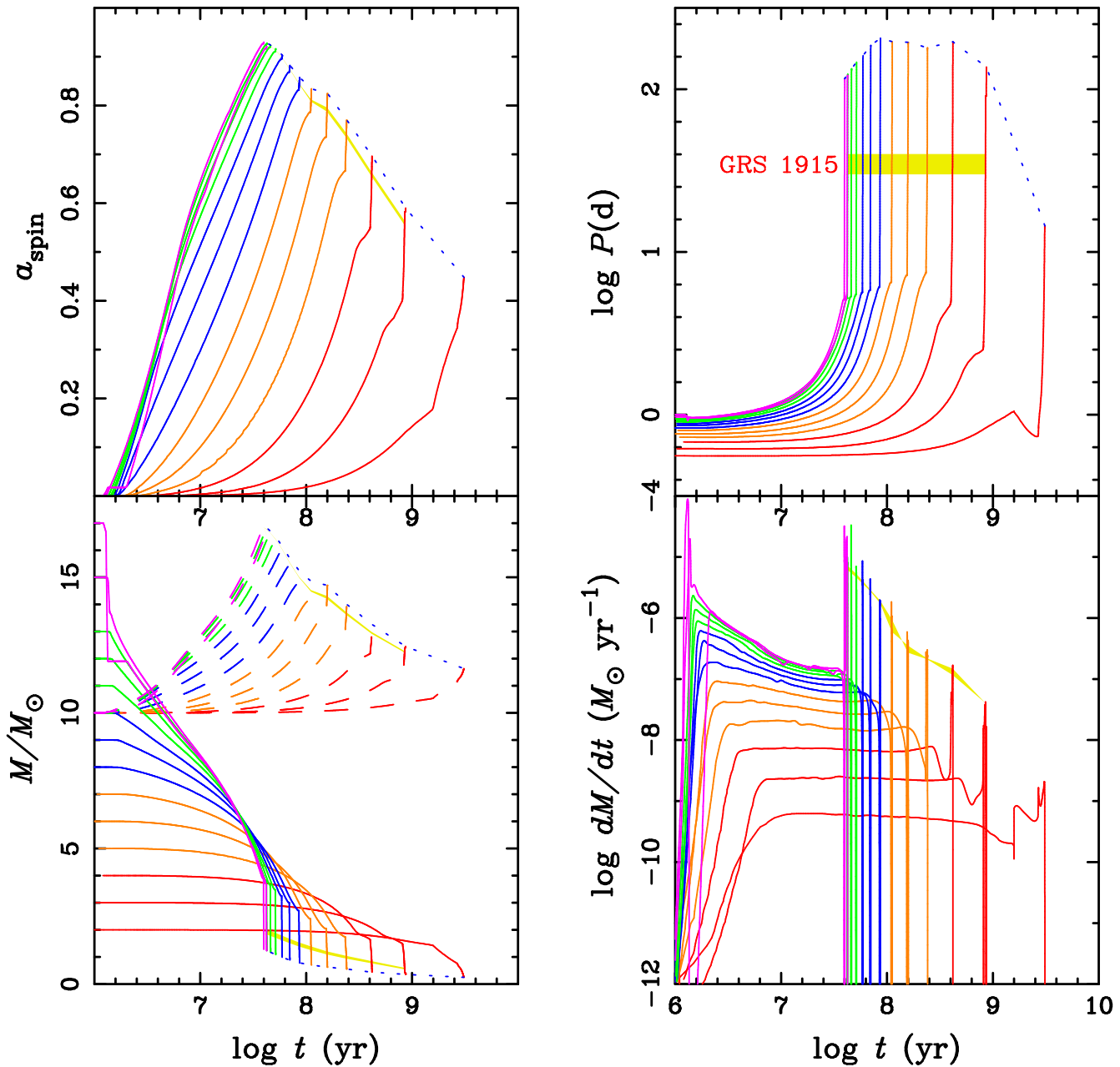}}
\caption{Selected properties of the
evolutionary sequences for black-hole binaries with initially
unevolved secondaries of 2 to 17\Msun\ (see Table~1) as a function of
time since the beginning of mass transfer (with arbitrary offset). The
initial mass of the black hole is 10\Msun\ in all sequences. Top left
panel: black-hole spin parameter ($J/M^2$); top right panel: orbital
period; bottom left panel: black-hole mass (dashed curves) and
secondary mass (solid curves); bottom right panel: mass-transfer rate.
The shaded regions in each panel indicate the period range of 30 to
40\,d {(similar to the orbital period of GRS 1915+105 with P$_{\rm
orb}$ = 33.5\,d).}}\label{Fig4}
\end{figure*}

All our calculations were performed with a standard Henyey-type stellar
evolution code (Kippenhahn, Weigert \& Hofmeister 1967) which we have used
in various similar investigations in the past and which is described
in detail in Podsiadlowski, Rappaport \& Pfahl (2002b; hereafter PRP).
It uses up-to-date stellar OPAL opacities
(Rogers \& Iglesias 1992), complemented
with those by Alexander \& Ferguson (1994) at low temperatures. In all
calculations we assumed an initial solar composition with $X=0.70$ and
$Z=0.02$, took a mixing-length parameter $\alpha=2.0$, and included
0.25 pressure scale heights of convective overshooting from the core
(Schr\"oder, Pols \& Eggleton 1997; Pols et al.\ 1997).

Since the accreting object is a black hole, we need to follow the change
in the accretion efficiency and the change in the spin angular momentum
of the black hole as its mass increases. In our calculations we assume
for simplicity that the black hole is initially non-rotating.
If we further assume that the efficiency, $\eta$, with which the
black hole radiates is determined by the last stable particle orbit,
then the black-hole luminosity can be written as
\begin{equation}
 L= \eta\,\dot{M}_{\rm acc}c^2,\label{leta}
\end{equation}
where $\dot{M}_{\rm acc}$ is the black-hole mass-accretion rate (as
measured by an observer at infinity),
$c$ is the speed of light and $\eta$ is approximately given by
\begin{equation}
\eta = 1-\sqrt{1-\left({M_{\rm BH}\over 3M_{\rm BH}^0}\right)^2}.
\label{eta}
\end{equation}
for $M_{\rm BH} < \sqrt{6} M_{\rm BH}^0$. Here, $M_{\rm BH}^0$ and
$M_{\rm BH}$ are the initial and the present gravitating mass-energy
of the black hole, respectively (Bardeen 1970; see also King \& Kolb
1999). In none of our evolutionary sequences does $M_{\rm BH}$ exceed
$\sqrt{6} M_{\rm BH}^0$. Over this interval $\eta$ ranges from $\sim
0.06-0.42$ (but see footnote 2).  If the black hole were born with
significant rotation (as argued e.g. by LBW), all of these
expressions would have to be modified accordingly.

This luminosity needs to be compared to the Eddington luminosity at which
the radiation pressure force balances gravity
\begin{equation}
L_{\rm edd} = {4\pi G M_{\rm BH} c\over \kappa},
\end{equation}
where $G$ is the gravitational constant and $\kappa$ is the opacity
assumed to be due to pure electron scattering, i.e.\
$\kappa=0.2\, (1+X)\, {\rm cm}^{2}\,{\rm g}^{-1}$ for a composition with
hydrogen mass fraction $X$ (e.g. Kippenhahn \& Weigert 1990).
Equating $L_{\rm edd}$ to $L$ in equation~(\ref{leta}) then defines the
Eddington mass-accretion rate, i.e. the maximum accretion rate
at which gravity can overcome radiation pressure (for spherical accretion):
\begin{equation}
\dot{M}_{\rm edd} = {4\pi G M_{\rm BH}\over \kappa c\eta}
\end{equation}
\begin{equation}
\hfill\simeq 2.6\times 10^{-7}\Msyr\left({M_{\rm BH}\over 10\Msun}\right)\,
\left({\eta\over 0.1}\right)^{-1}\,\left({1+X\over 1.7}\right)^{-1}
\label{edd}
\end{equation}
In most calculations we assume that any mass transferred in excess
of the Eddington accretion rate is lost from the system, carrying with
it the same specific angular momentum as the orbiting black hole, while
the rest of the mass, reduced by the fractional rest mass energy lost in
the radiation, is accreted by the black hole.

As the black hole accretes mass and angular momentum, its
spin parameter, $a\equiv J/M^2$, increases according to
\begin{equation}
a=\left({2\over 3}\right)^{1/2}\,{M_{\rm BH}^0\over M_{\rm BH}}\,
\left[4-\left(18 \left({M_{\rm BH}^0\over M_{\rm BH}}\right)^2
-2\right)^{1/2}\right]\label{spin}
\end{equation}
for $M_{\rm BH} < \sqrt{6} M_{\rm BH}^0$ (see e.g. Bardeen 1970;
Thorne 1974; King \& Kolb 1999)\footnote{For a maximally spinning
Kerr black hole $a$ nominally approaches unity, but is probably
limited by a counteracting torque due to disk radiation swallowed by the
black hole to be $\sim 0.998$ (Thorne 1974). It also limits the
efficiency $\eta$ in equation~(\ref{eta}) to $\sim 0.30$.}.

Our calculations include, as a default, orbital angular momentum
losses via magnetic braking if the secondary has a convective envelope
(see e.g. Verbunt \& Zwaan 1981; Rappaport, Verbunt \& Joss 1983),
and by gravitational radiation; these affects are only important in
some of the calculations with relatively low-mass donor stars in \S~3.2
(also see \S~4.4). In most cases mass transfer occurs either on a
thermal timescale or is driven by the nuclear evolution of the secondary.

\noindent
\subsection{Results of the Binary Evolution Calculations}

\begin{figure*}
\centerline{\psfig{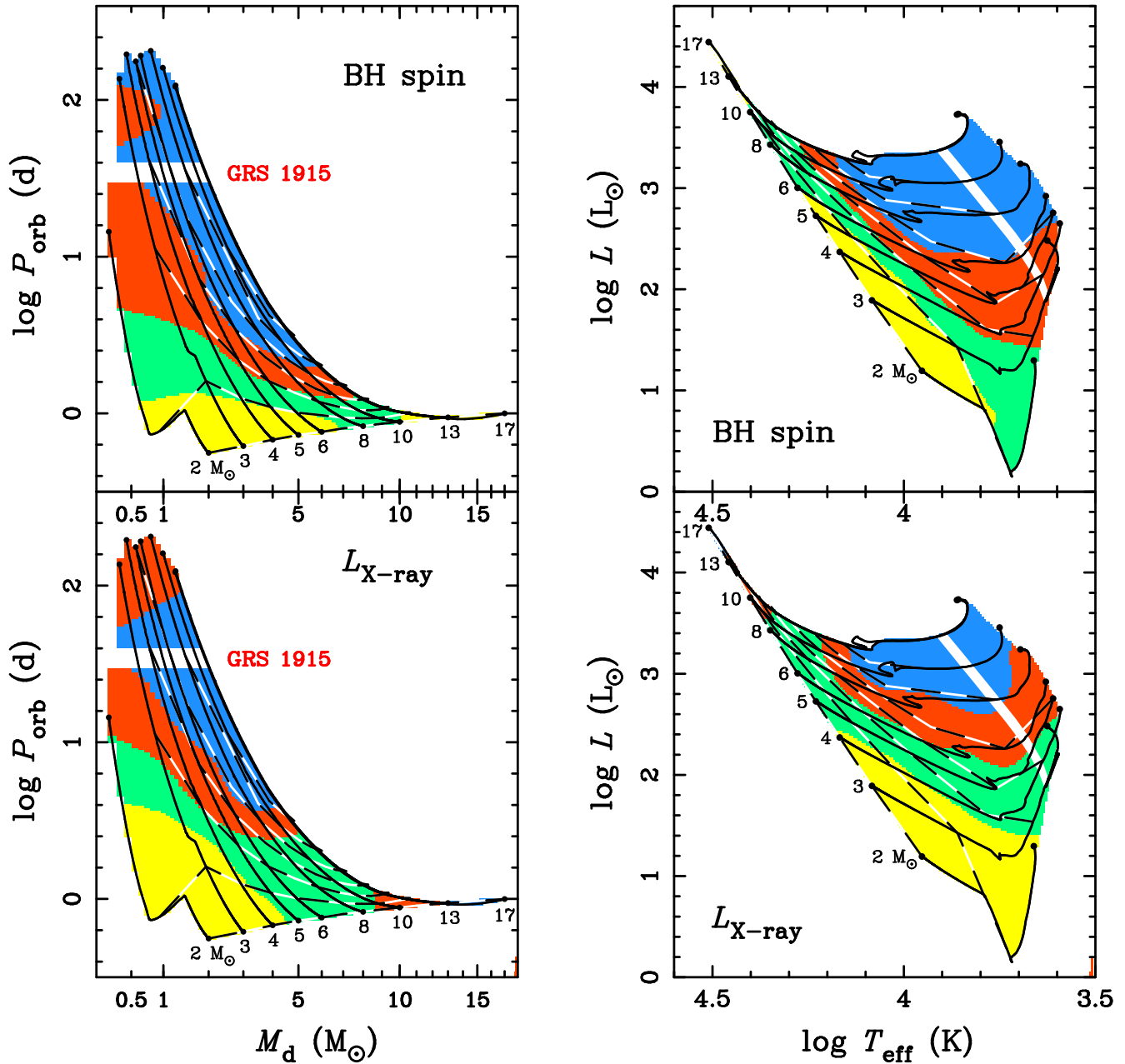}}
\caption{Shaded contours of black-hole spin parameter
(${J/M^2}$) (top panels) and {\em potential\/} X-ray luminosity (bottom
panels) (i.e.\ the luminosity assuming that all the mass transferred
is accreted by the black hole) in the P$_{\rm orb}$\,--\,M$_{\rm d}$
(orbital period -- secondary mass) plane (left panels) and in the H-R
diagram (right panels).  Shading for the black-hole spin (from light
to dark): 0\,--\,0.25; 0.25\,--\,0.50; 0.50\,--\,0.75; 0.75\,--\,1.00.
Shading for the {\em potential\/} X-ray luminosity (from light to dark):
transient X-ray sources; steady sources with L$_{\rm X}<$ 10$^{39}
$\,ergs\,/s); 10$^{39}<$ L$_{\rm X}<$ 10$^{40}$\,ergs\,/s; L$_{\rm
X}>$ 10$^{40}$\,ergs\,/s.  The solid curves show the evolutionary
tracks for the initially unevolved secondaries in Table~1 and
Figure~\ref{Fig4} with the initial masses as indicated.  The dashed
curves are contours of constant black-hole mass (10, 11, 12, 13, 14, 15,
16\Msun\ from left to right [right panels], bottom to top [left
panels]). The white strips indicate the position of systems with
orbital periods between 30 and 40\,d, i.e. have an orbital period
similar to GRS 1915+105.}\label{Fig5}
\end{figure*}

To illustrate the evolution of black-hole binaries we first performed
a sequence of models where we varied the mass of the secondary from 2
to 17\Msun. In these calculations the secondary was an initially
unevolved (i.e. zero-age main-sequence) star and the black-hole had
an initial mass of 10\Msun. Furthermore, we assumed that the
mass-accretion rate onto the black hole was Eddington-limited
(eq.~\ref{edd}).  Some of the main results of these sequences are
shown in Figures~\ref{Fig4} and \ref{Fig5} and in the top part of
Table~1. All calculations were terminated either when the secondary
became degenerate or when it became detached during helium core
burning. In some cases, the secondary would have filled its Roche lobe
again after helium core burning and ascended the asymptotic giant
branch. In this relatively short-lived phase the properties of the
systems would be similar to those on the ascent of the first giant branch
and the orbital period would continue to increase somewhat (but
typically by less than a factor of 2). In all cases with initially
unevolved secondaries, the secondaries end their evolution as white
dwarfs rather than in a supernova.

\begin{table*}
\hspace{-4.5cm}{\psfig{file=table1.epsi,width=0.7\textwidth}}
\end{table*}

The main behaviour of these binary sequences is not difficult to
understand; it is mainly determined by the initial mass ratio of the
components ($q\equiv M_{\rm d}/M_{\rm BH}$; see e.g. Ritter 1996; Kalogera
\& Webbink 1996; PRP for detailed recent discussions). Generally the
mass-transfer rates (bottom right panel in Fig.~\ref{Fig4}) at any
point in the evolution are higher for higher initial secondary masses
(mass ratios). For relatively low-mass secondaries, the secondaries
always remain close to thermal equilibrium and mass transfer is
entirely driven by the nuclear evolution of the secondary. The
mass-transfer rate tends to dip near the end of the secondary's
main-sequence phase, where the secondary may even become detached
temporarily, and then increases sharply as the secondary ascends the
giant branch where the evolution is determined by the rate at which
the hydrogen-burning shell advances through the star.  For the more
massive secondaries ($q>1$), the mass-transfer timescale becomes
shorter than the thermal timescale of the secondary and mass transfer
occurs initially on the thermal timescale of the secondary's
envelope. For the most massive secondaries this leads to a sharp
initial spike in the mass-transfer rate, reaching a peak of
$\dot{M}\sim 10^{-4}\Msyr$. Note that after this initial high
mass-transfer rate, the evolution of all sequences with $M\ga 12\Msun$
becomes essentially uniform and independent of the initial mass of the
secondary\footnote{This is only true for secondaries that are
initially relatively unevolved which after the initial high
mass-transfer phase behave like lower-mass secondaries. If the
secondaries have already established a very non-uniform chemical
composition profile, the subsequent evolution can be drastically
altered (see e.g. Podsiadlowski \& Rappaport 2000).}.

For secondaries with initial masses larger than $\sim 20\Msun$
(not shown in the figures; but see \S~4.3), the
initial mass-transfer rate would become so
high that mass transfer would become unstable (i.e. experience a
delayed dynamical instability; Hjellming \& Webbink 1987; PRP)
and the secondary would then be likely to engulf the black hole
leading to a second common-envelope phase and the
spiral-in of the black hole in this envelope\footnote{\label{foot}In
the calculation with an initially unevolved 20\Msun\ star, the
secondary never overfilled its Roche lobe by more than 14 per cent
(for a short period of time). It is not entirely clear whether this
necessarily leads to a spiral-in phase, in particular for stars with
radiative envelopes (Podsiadlowski 2001). If it does not, and the
system survives the initial phase of high mass transfer, the secondary
becomes detached after it has reestablished thermal equilibrium
(similar to Fig.~7 in PRP);
but the secondary now has a mass of only 4\Msun\ and the
subsequent evolution will mimic the evolution of a system with a
4\Msun\ secondary. It starts to fill its Roche lobe shortly after the
hydrogen core-burning phase (i.e. experience early case B mass
transfer) and has parameters at an orbital period of 33.5\,d
consistent with those of GRS 1915+105 (see \S~4.1).}.  The further
evolution in this case is presently rather uncertain; the black hole
will most likely settle at the center of the secondary destroying its
core or merging with it.  The ultimate product will be a black hole
with a possibly much larger mass than the initial mass if it is able
to accrete a substantial fraction of the secondary (e.g. if accretion
occurs in the super-critical regime where radiation can be trapped in
the flow; Houk \& Chevalier 1991; Chevalier 1993). In addition, the
black-hole may be surrounded by planet-mass objects or one or more
low-mass stars that are likely to form by gravitational instabilities
in the centrifugally supported disk left over from the collapsing
envelope of the secondary (for more details of such a scenario see
Podsiadlowski et al.\ 1995).  If the black hole is indeed orbited by a
low-mass stellar companion, the system may later appear again as an
X-ray binary, but would now be classified as a low-mass black-hole
binary. Note, however, that the secondary should be chemically
anomalous for its mass since it is likely to contain matter that
underwent nuclear processing in a massive star (i.e.\ show evidence
for CNO processing).

Returning to Figure~\ref{Fig4}, the left bottom panel shows that
the black-hole mass (dashed curves) can increase
substantially in these sequences, by up to $\sim 7\Msun$
(also see Table~1), even though we assumed that mass
accretion onto the black hole was Eddington-limited. Most of this
accretion takes place when the secondary is still burning hydrogen in
the core since this phase lasts much longer than the subsequent giant
phase and since the mass-transfer rates on the giant branch are
generally much higher, often significantly super-Eddington. This has
the consequence that a larger fraction of the transferred mass is lost
from the system.

Our finding that the black hole can accrete a fairly significant
amount of mass is in apparent conflict with the results of a related
study by King \& Kolb (1999) who found a much more moderate possible
increase of the black-hole mass ($\la 2.5\Msun$) by estimating the
maximum amount that can be accreted as the product of the Eddington
accretion rate and the evolutionary timescale of the secondary. The
differences in these results can be attributed to two factors. First
King \& Kolb (1999) used a characteristic value for the Eddington
accretion rate ($10^{-7}\Msyr$ for a 10\Msun\ black hole) that is
substantially lower (up to a factor of $\sim 4$) than the value given
by equation~(\ref{edd}) in the early phase of the evolution when the
black-hole radiation efficiency is low ($\eta \simeq 0.06$). A second
factor is that, in particular for the more massive secondaries, the
characteristic evolutionary timescale increases as the mass of the
secondary decreases and as the secondary behaves like a less massive
star.  This increases the evolutionary timescale by between a factor
of $\sim 2.5$ for the least massive secondaries to $\sim 4$ for the
more massive secondaries in our calculations (a factor of 100 in
the 20\Msun\ calculation).

However, all of the calculations presented so far assume that the
secondary is initially unevolved. As the results in \S~2 show, it is
much more likely that the secondary was at least somewhat evolved at
the beginning of mass transfer. Since this shortens the remaining
evolutionary time in the hydrogen core-burning phase, it reduces the
amount of matter that can be accreted by the black hole.  To
illustrate this we have performed four evolutionary sequences for a
secondary with an initial mass of 8\Msun\ at different evolutionary
stages (see `Evolved Sequences' in Table~1). In these calculations the
secondaries had a hydrogen mass fraction in the core at the beginning
of mass transfer of $X= 0.50$, 0.30, 0.10 in the first three
sequences, respectively, while in the fourth sequence the secondary
already had developed a hydrogen-exhausted core of 0.7\Msun\
(i.e. experienced so-called early case B mass transfer). As expected,
the final black-hole mass decreases systematically from 15.2\Msun\ for
the initially unevolved secondary to only 11.1\Msun\ for the secondary
near the end of the main-sequence phase. In the calculation where mass
transfer starts after the main-sequence phase of the secondary, the
black hole accretes only $\sim 0.05\Msun$.  Nevertheless, even for an
initially moderately evolved secondary, the black-hole mass can still
increase quite substantially (by $\sim 4\Msun$).

As the black hole accretes matter from the last stable orbit, it
also accretes angular momentum and is spun up in the process.
The top left panel in Figure~\ref{Fig4} shows the time evolution
of the black-hole spin parameter $a$ (eq.~\ref{spin}). Even if
the black hole was completely non-rotating initially (as we assumed
in our sequences) and the accretion rate is Eddington limited,
the black hole can be spun up substantially to a spin parameter
$a\sim 0.9$, where for a maximally rotating Kerr black hole
$a=0.998$ (Thorne 1974).

In Figure~\ref{Fig5} we show the distribution (indicated by
shading) of the spin parameter (top panels) and the
{\em potential\/} X-ray luminosity, defined below, (bottom
panels) for selected evolutionary sequences of the secondary
both in the orbital period -- secondary mass plane (left panels),
and in the \mbox{H-R} diagram (right panels). In these figures, the solid
curves represent selected evolutionary sequences with initially
unevolved secondaries (as indicated). Note in particular how the
evolutionary tracks for the most massive secondaries all converge and
how the effective temperature in the red-giant phase increases
systematically with initial secondary mass. The latter is in part
caused by the much lower hydrogen abundance in the envelope of the
initially more massive secondaries (see Table~1). The dashed curves in
the panels indicate the black-hole mass (from 10 to 16 \Msun).

In the bottom panels of Figure~\ref{Fig5} we have defined a '{\em
potential\/} X-ray luminosity' as the accretion luminosity from the
black hole assuming that all the matter that is transferred from the
secondary is accreted by the black hole radiating at the appropriate
efficiency $\eta$ (eq.~\ref{eta}), i.e. assuming that accretion is not
Eddington limited (see e.g. Begelman 2002). Because of the high
mass-transfer rate, in particular for the more massive systems, this
{\em potential\/} X-ray luminosity can be as high $\sim 10^{41}\ergss$
and systems can spend a large fraction of their X-ray-active lifetime
at these high mass-transfer rates (see the X-ray lifetimes in Table~1
and the further discussion in \S~4.2).  Figure~\ref{Fig5} also shows
where systems would be expected to be black-hole transients (light
shading). To decide whether a system exhibits transient behaviour, we
utilized an expression very similar to equation (A5) of Vrtilek et
al.\ (1990) for determining the outer disk temperature at $r_d$.  If
$T_e(r_d)$ is found to be $\ga 6500$ K, we take the disk to be
ionized, and therefore {\em not} subject to the standard
thermal-ionization disk instability (Cannizzo \& Wheeler 1984; van
Paradijs 1996; King, Kolb \& Szuskiewicz 1997; Lasota 2002).  As
expected from the behaviour of the mass-transfer rates
(Fig.~\ref{Fig4}), systems with relatively low-mass secondaries tend
to be transient systems, and the {\em potential\/} X-ray luminosities
increase systematically with initial secondary mass.

When interpreting these results, several caveats are in order.  The
actual binary evolution calculations leading to the results discussed
above assumed that any mass transferred in excess of the Eddington
rate was lost from the system; this, in turn, somewhat affects the evolution of
the orbit and the mass-transfer rate itself. In order to estimate how
sensitive our results for the {\em potential\/} luminosity are to the
assumption of Eddington-limited accretion, we performed two sequences
where we assumed that accretion onto the black hole was not Eddington
limited for secondaries with an initial mass of 8 and 15\Msun,
respectively (see `Non-Eddington limited sequences' in Table~1). The
8\Msun\ sequence is only moderately affected since the mass-transfer
rate is sub-Eddington for most of the evolution, while in the 15\Msun\
sequence the black-hole mass grows significantly larger, as expected,
than in the Eddington-limited case. But even for the more massive
secondary, the mass-transfer rates are typically within a factor of
two at the same orbital period for the Eddington-limited and the
non-limited case. We therefore conclude that the inferred {\em
potential\/} X-ray luminosities are correct as computed to within a
factor of a few.  Since the Eddington luminosity in our systems with
the highest mass black holes ($\sim 17\Msun$) is $4 \times
10^{39}\ergss$, an observed X-ray luminosity as high as
$10^{40}\ergss$ would require only a modest super-Eddington mass
accretion rate of a factor of a few $\dot{M}_{\rm edd}$. Even for
systems with {\em potential\/} X-ray luminosities as high as
$10^{41}\ergss$, mass accretion has to exceed the Eddington accretion
rate by typically less than a factor $\sim 20$. These super-Eddington
accretion rates may be significantly reduced if beaming of the X-ray
flux is important in these systems (King et al.\ 2001). Begelman
(2002) has recently reexamined the problem of super-Eddington
accretion and concluded that in radiation-pressure dominated accretion
discs super-Eddington accretion rates of a factor of 10 to 100 can be
achieved, due to the existence of a photon-bubble instability in
magnetically constrained plasmas.  In this context, it is worth
pointing out that a number of X-ray binaries containing neutron stars
are known to radiate substantially above the Eddington limit (SMC X-1,
Levine et al.\ 1993; LMC X-4, Levine et al.\ 1991) by factors of up to
$\sim 5$. It is believed that this is a consequence of the fact that
the accretion onto the poles of the neutron star is funneled through a
strong magnetic field, a process that is not directly applicable to
black-hole systems. Nevertheless, it demonstrates that super-Eddington
X-ray binaries exist in Nature (also see \S~4.1).

It is also worth pointing out that the mass-transfer rates obtained
from our calculations are secular mass-transfer rates, i.e. represent
an average over timescales much longer than the lifetime of X-ray
astronomy. It is quite plausible that the mass-transfer rates
fluctuate substantially about the secular mean even in systems that
are not considered `transients' according to the disc-instability
model. GRS 1915+105 may present an example for this. It became
an X-ray source in 1992 (Castro-Tirado, Brandt \& Lund 1992)
and has been a relatively steady source ever since (Sazonov et al.\ 1994;
Greiner, Morgan \& Remillard 1997; http://xte.mit.edu).
Its behaviour is very different from the normal behaviour
of soft X-ray transients and the system could probably be better
classified as a semi-persistent source.

\section{Applications}

\begin{figure*}
\centerline{\psfig{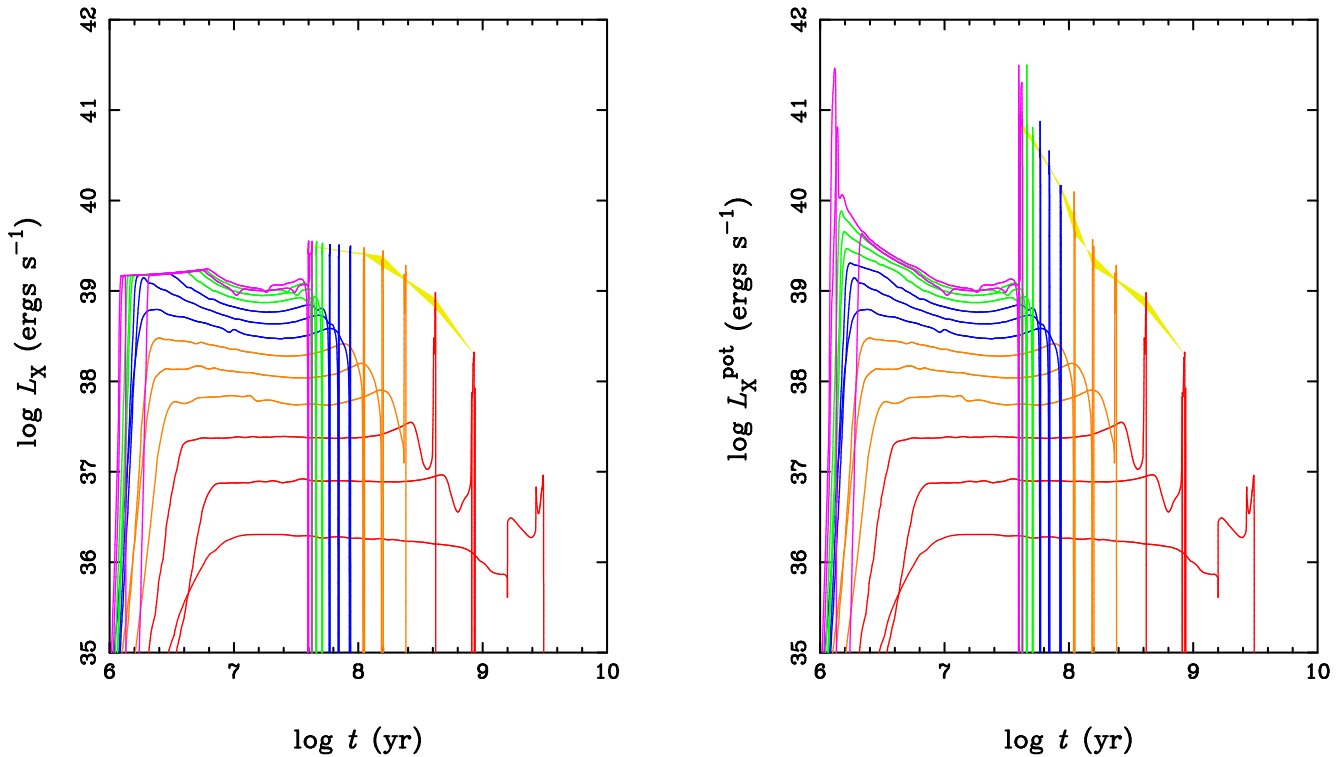}}
\caption{X-ray luminosity, assuming Eddington-limited accretion, (left)
and {\em potential\/} X-ray luminosity, assuming non-Eddington-limited
accretion, (right) for the binary sequences with unevolved secondaries
(see Fig.~\ref{Fig4}) as a function of time since the beginning of
mass transfer. The shaded regions in each panel indicate the period
range of 30 to 40\,d (similar to the orbital period of GRS 1915+105
with P$_{\rm orb}$ = 33.5\,d).}\label{Fig6}
\end{figure*}

\subsection{GRS 1915+105}

One of the best-studied black-hole binaries in the Galaxy is the
microquasar GRS 1915+105 (see e.g. Castro-Tirado, Brandt \& Lund 1992;
Greiner, Morgan \& Remillard 1996; Mirabel \& Rodr\'\i guez 1994).
Greiner, Cuby \& McCaughrean (2001) recently determined the orbital
period of the system as $33.5\pm 1.5\,$d and obtained the black-hole
mass function ($f(M)= 9.5 \pm 3.0\Msun$). Based on their analysis,
they find a black-hole mass of $14\pm 4\Msun$, substantially more
massive than the masses inferred for the majority of black-hole
transients (see e.g. Table 1 in LBW and references therein; Orosz et
al.\ 2002). In Figures~\ref{Fig4} and \ref{Fig5}, we indicated the
period range of 30 to 40\,d, i.e. close to the period of GRS 1915+105,
in all panels and in Table~1 we give some of the key system parameters
of all evolutionary sequences at an orbital period of 33.5\,d.  The
sequence with an initially unevolved secondary of 2\Msun\ never
reaches an orbital period of 33.5\,d. The reason is that in this case
(unlike the other cases), magnetic braking becomes important, indeed
the dominant mass-transfer driving mechanism, at the end of the
donor's main-sequence phase; this causes a peak in $\dot{M}$ and a
temporary shrinking of the orbit (see Fig.~4).  We therefore added
another sequence with an initial secondary of 2\Msun\ which has
already evolved off the main sequence at the beginning of mass
transfer (i.e. experienced case B mass transfer). Since the evolution
of low-mass giants is entirely determined by the evolution of the core
mass (and not the total mass), this sequence can also be considered
representative for systems with low-mass secondaries in general (as
assumed e.g. by Belczynski \& Bulik 2002; Vilhu 2002 in their
modelling of GRS 1915+105).

As Table~1 shows, the masses of both components and the mass-transfer
rate at an orbital period of 33.5\,d increase systematically with the
mass of the secondary, reaching a maximum for an initial secondary of
$\sim 15\Msun$. The same applies to the effective temperature and the
luminosity. Greiner et al.\ (2001) have estimated the spectral type of
the secondary as K-M III. Since a K0 III star has an effective
temperature of $\sim 4800\,$K (e.g. Strai\v{z}ys \& Kuriliene 1981),
we can use this as an additional constraint to limit the possible
evolutionary histories of GRS 1915+105. Inspection of Table~1 then
suggests that acceptable models for GRS 1915+105 can have initial
secondary masses as high as 6\Msun. Indeed the model parameters for
GRS 1915+105 in the 6\Msun\ sequence are very close to the system
parameters deduced by Greiner et al.\ (2001; $M_{\rm BH} \sim 14 \Msun$,
$M_{\rm d}\sim 1.2\Msun$), for an assumed system inclination angle of $\sim
70^{\circ}$, as determined from the orientation of the jets (Mirabel \&
Rodr\'\i guez 1994). We note that in principle one could use
the `transient' nature of GRS 1915+105 to further constrain the
evolutionary past of the system. However, considering the unusual
X-ray behaviour of GRS 1915+105, which is very different from the 
predictions of the simple disc-instability model, this is probably 
not advisable (see the discussion at the end of \S~3.2).

Our calculations have several important implications for GRS 1915+105.
First, they show that the mass of the black hole may have increased
significantly, by up to $\sim 4\Msun$, even if the mass accretion rate is,
on average, Eddington limited (also see LBW who obtained
a similar estimate). Hence the present mass of the black hole is
not a good indicator of the initial black-hole mass, and any analysis
of the implications of observed black-hole masses for the evolution
of the black-hole progenitor has to take this into account. We have
calculated some additional evolutionary sequences starting with a
lower black-hole mass of 7\Msun\ (similar to the masses found in other
black-hole binaries) and obtained acceptable models for GRS 1915+105
with black-hole masses as high as 11\Msun.

Second, since the black hole may have accreted a substantial amount of
matter, it may also have been spun up significantly and may have acquired
a spin parameter as high as $a\sim 0.8$ (assuming an initially
non-rotating black hole). This may be important for modelling the jets
and the emission from the inner parts of the accretion disc
(e.g. Zhang, Cui \& Chen 1997; Makishima et al.\ 2000).

Third, the secular mass-transfer rate can be as high as $\sim 3\times
10^{-7}\Msyr$, implying an X-ray luminosity as high as $\sim 2\times
10^{39}\ergss$.  This is a factor of a few lower than the peak X-ray
luminosity of $7\times 10^{39}\ergss$, determined for GRS 1915+105 by
Greiner, Morgan \& Remillard (1996). This implies that only a moderate
amount of super-Eddington accretion is required to explain the
observed peak luminosity in our models to explain the observed
luminosity.

Fourth, our models predict that the surface abundance of the secondary
should be substantially enhanced in helium and could show CNO
abundance ratios that are close to the equilibrium ratios for the CNO
cycles (in particular for the more massive secondaries).  This
provides a potentially powerful test that may help to further
constrain the nature and the initial mass of the secondary.

\subsection{Ultraluminous X-ray sources}

Ultraluminous X-ray sources (ULX) are luminous X-ray sources outside
the nuclei of external galaxies, typically defined to have an X-ray
luminosity larger than $10^{39}\ergss$. They were originally
discovered by {\sl Einstein} (Fabbiano 1989) and have been found in
large numbers by {\sl ROSAT} and most recently {\sl Chandra} (Colbert
\& Mushotzky 1999; Roberts \& Warwick 2000; Colbert \& Ptak
2002; Lira, Johnson \& Lawrence 2002; Jeltema et al.\ 2002).  While
the physical nature of these sources has remained unclear, and indeed
they probably form a heterogeneous class of systems (Prestwich et al.\
2002; Roberts et al.\ 2002), a plausible scenario is that they
constitute the luminous tail of the stellar-mass black-hole binary
distribution (for recent discussions see Colbert \& Mushotzky 1999;
King et al.\ 2001; King 2002; Roberts et al.\ 2002). With an inferred
peak X-ray luminosity of $7\times 10^{39}\ergss$ (Greiner et al.\
1996), GRS 1915+105 certainly classifies as a typical ULX (King et
al.\ 2001; Mirabel \& Rodr\'\i guez 1999). The connection of ULXs with
`normal' Galactic black-hole binaries has been strengthened by the
determination of the black-hole mass in GRS 1915+105 (Greiner et al.\
2001), which proved that ULXs may contain typical stellar-mass black
holes rather than a previously unknown class of intermediate-mass
black holes of $10^2\,$--$\,10^4\Msun$ (as suggested by Colbert \&
Mushotzky 1999).

The calculations presented in this paper strongly support the
connection of ULXs with black-hole binaries, in particular those with
intermediate-/high-mass secondaries. Figure~\ref{Fig6} shows both the
X-ray luminosity (assuming Eddington-limited accretion) and the {\em
potential\/} X-ray luminosity (assuming non-Eddington-limited
accretion) of the binary sequences. While the Eddington-limited
luminosity reaches a maximum of $\sim 4\times 10^{39}\ergss$, the {\em
potential\/} X-ray luminosity can be as high as $\sim 3\times
10^{41}\ergss$ and could be even higher if the mass-transfer rate is
variable (consistent with observations; see e.g. Lira et al.\ 2002 and
the discussion at the end of \S~3.2). Our calculations therefore show
that the mass-transfer rates in these sequences are high enough to
provide a potential power source for ULXs. It requires only that the
majority of ULXs have to radiate at a moderately super-Eddington
luminosity, as is actually observed in GRS 1915+105. Even for the most
luminous observed ULXs, the luminosity has to exceed the Eddington
luminosity by a factor of $\la 20$, which may not pose a serious
problem in radiation-pressure dominated magnetic discs (Begelman 2002). This
factor may be further reduced significantly if the radiation is beamed
(King et al.\ 2001). Figure~\ref{Fig5} shows where the most luminous
systems (dark shading) are expected to lie in the H-R diagram of the
secondary and in the orbital period -- secondary mass plane.  They
indicate that black-hole binaries are most likely to appear as ULXs
when they are giants or supergiants where mass transfer is driven by
the nuclear evolution of the secondary, a phase that can last up to
several $10^7\yr$ (also see Fig.~\ref{Fig6} and Table~1).

If ULXs in external galaxies are associated with
intermediate-/high-mass black-hole binaries, they should be
preferentially found near star-forming regions. Indeed there is some
evidence that many ULXs are found in regions of active star formation,
starburst galaxies and interacting galaxies (Lira et al.\ 2002;
Roberts et al.\ 2002; Terashima \& Wilson 2002; Zezas et al.\ 2002).
On the other hand, Colbert \& Ptak (2002) found that the number of
ULXs per galaxy is actually higher in elliptical than in
non-elliptical galaxies and that there is a population of ULXs in the
halos of elliptical galaxies, where recent star formation is not
expected to have occurred. This may suggest that our models are not
directly applicable to these ULXs, unless these galaxies experienced
some relatively recent star formation (e.g. as a result of some
previously unrecognized merger activity). King (2002) has recently
proposed that there are two classes of ULXs: relatively massive
systems associated with young stellar populations and low-mass systems
found in old populations. Indeed there is some evidence that a large
fraction of ULXs in elliptical galaxies are located in globular
clusters (see e.g. Angelini, Loewenstein \& Mushotzky 2001; Kundu,
Maccarone \& Zepf 2002; White, Sarazin \& Kulkarni 2002; Jeltema
et al.\ 2002). These could be relatively long-period systems with
giant donors (at the current epoch) that formed dynamically in the dense
cluster cores by processes not considered in the present study.

In principle we could combine the results of our evolutionary
calculations with the BPS model in \S~2 and predict the properties of
black-hole binaries in a typical galaxy (e.g. the X-ray luminosity
function; see Roberts \& Warwick 2000). However, there are a large
number of uncertainties involved, including the histories of stellar and
globular-cluster formation in galaxies (especially for ellipticals), as
well as those associated with the modelling of the evolution that leads to
the formation of a black-hole binary and the problem of how to relate
secular mass-transfer rates to observable X-ray luminosities.  In view
of these, we cannot yet hope to reliably compute a population synthesis of
black-hole binaries in external galaxies, and therefore little of substance
is to be gained from such an exercise. However, putting aside some of these
issues, e.g. the question of super-Eddington accretion, and taking
the {\em potential\/} X-ray luminosity as a measure of the X-ray luminosity,
one can obtain a qualitative idea of the relative number of systems by
considering how much time systems spend at various mass-transfer rates.

In Table~1 we list the X-ray lifetimes for transient phases and
various ranges of {\em potential\/} X-ray luminosity in our binary
sequences. Systems with relatively low-mass secondaries are expected
to be transients for most of their evolution, while the systems with
the most massive secondaries spend a large fraction of their X-ray
lifetimes as ULXs. Considering that our standard BPS model predicts a
fairly uniform distribution of secondary masses, at least above some
characteristic minimum mass (see Figure~\ref{Fig2}), and taking the
11\Msun\ donor sequence as typical, we then estimate that
some 10 per cent of black-hole binaries should be ULXs,
and some 10 per cent of these have
{\em potential\/} X-ray luminosities above $10^{40}\ergss$. This would
appear to be in qualitative agreement with the luminosity distribution
of Roberts \& Warwick (2000) who found that the number of sources in
each decade of X-ray luminosity decreases according to $L_{\rm
X}^{-0.8}$. Estimates of the absolute numbers of such binaries
residing in a typical Milky-Way type galaxy can be made by taking the
formation rates from Figure~\ref{Fig3} for black-hole binaries with
different initial donor masses and multiplying them by the various
lifetimes given in Table 1. Such an exercise shows that for
conventional values of the $\lambda$ parameter near $\sim 0.5$ the
predicted numbers of ULXs is substantially in excess of what is
observed.  However, for values of $\lambda$ closer to more realistic
values of $\sim 0.1$, the computed numbers of ULXs is quite
reasonable.

Finally we note a piece of observational evidence that suggests the
high luminosities inferred for some of the ULXs on the assumption of
roughly isotropic emission are, in fact, correct.  Pakull \& Mirioni
(2002) have studied very large ($\sim 300$ pc) and luminous ionization
nebulae surrounding about a dozen ULXs. They use the emission lines from
these nebulae as an interstellar medium calorimeter to infer the
long-term (e.g. $\sim 10^4$ yr; see e.g. Chiang \& Rappaport 1996)
mean X-ray luminosity of the underlying ULXs. From
the structure of the ionization nebulae they can also rule out in some cases
any extreme beaming of the radiation.  The results of this study seem to
point to good agreement between the X-ray luminosity determined from {\sl
Chandra} and {\sl ROSAT} and that found indirectly from studies of the
ionization nebulae.

\noindent
\subsection{Cygnus X-1}

\begin{figure*}
\centerline{\psfig{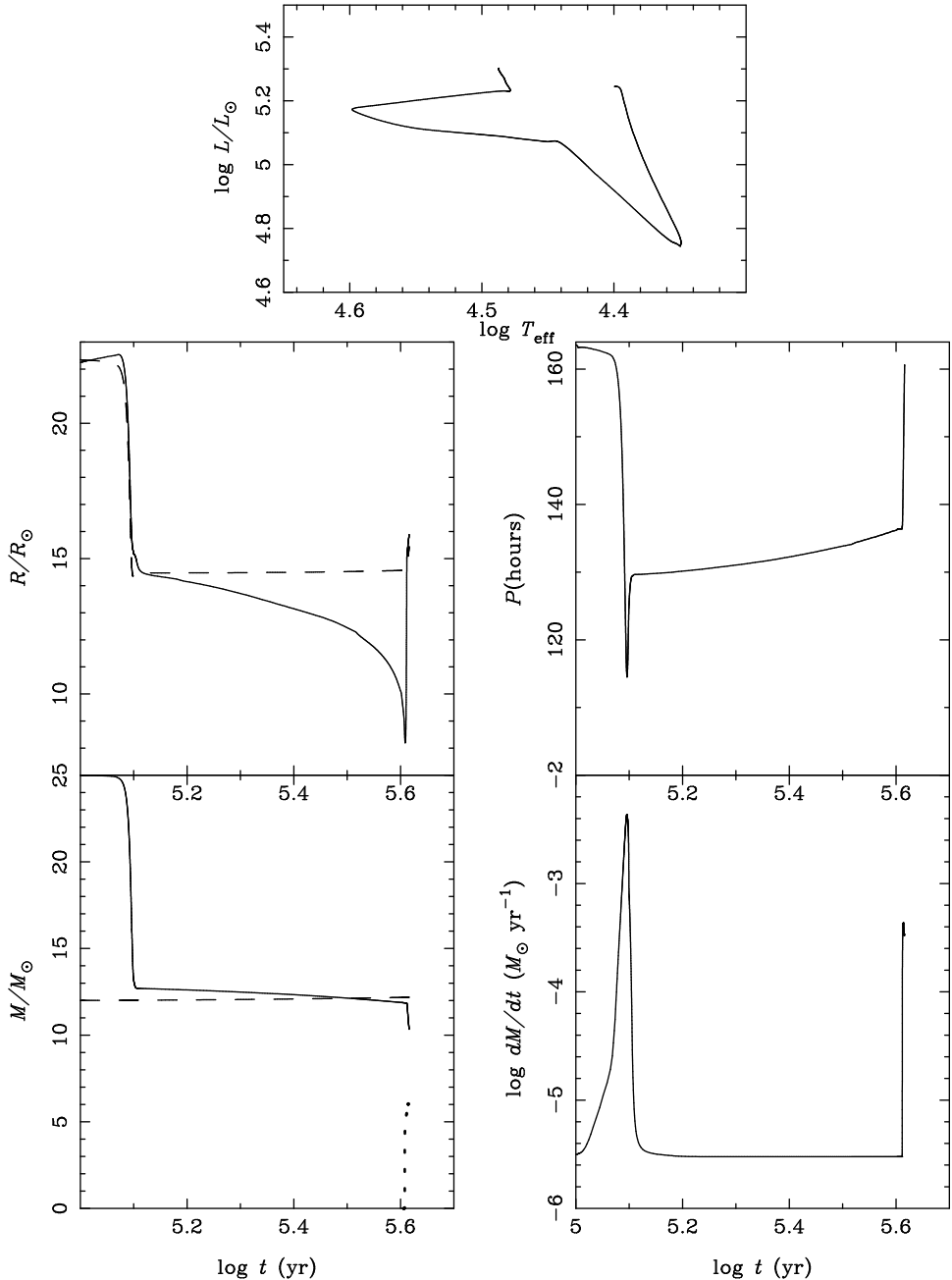}}
\caption{Evolutionary model for Cyg X-1. H-R diagram (top panel) and
key binary parameters as a function of time since the beginning of
mass transfer (with arbitrary offset). {\it Middle left:} Radius of
the secondary (solid curve) and Roche-lobe radius (dashed curve). {\it
Middle right:} Orbital period. {\it Bottom left:} Mass of the
secondary (solid curve), the black hole (dashed curve) and the mass of
the hydrogen-exhausted core (dotted curve). {\it Bottom right:}
Mass-loss rate of the secondary. The secondary initially has a mass of
25\Msun\ and has already exhausted most of its hydrogen in the core
(it has a core hydrogen mass fraction $X=0.054$). The initial mass of
the black hole is 12\Msun\ and barely changes during the
evolution. The calculation includes a constant stellar wind from the
secondary with a mass-loss rate of $3\times 10^{-6}\Msyr$, consistent
with the present observations of the secondary in Cyg
X-1. The black-hole accretion rate is always limited by the Eddington
accretion rate.}\label{Fig7}
\end{figure*}

One of the most famous black-hole binaries is Cyg X-1 with an orbital
period of 5.6\,d and a massive secondary (HDE 226868) of spectral type
O9.7 Iab (Walborn 1973; Gies \& Bolton 1986). It has a mass function
of $0.252\pm 0.010\Msun$ (Gies \& Bolton 1982) and used to be one of
the best early black-hole candidates before the identification of
low-mass black-hole transients with much larger mass functions.
However, the evolutionary state of the system has so far not been
properly established; in particular it is not clear how much mass the
donor star has already lost.

In Figure~\ref{Fig7} we present the results of a binary calculation
that may represent the evolution of Cyg X-1 and which illustrates
several characteristic properties of a massive black-hole binary.  In
this model, the initial masses of the black hole and the secondary
are 12 and 25\Msun, respectively, and the secondary starts to fill
its Roche lobe near the end of its main-sequences phase, when its
central hydrogen mass fraction has been reduced to 0.054. The orbital
period at this point is 6.8\,d. Unlike our previous calculations we
also included a stellar wind from the secondary of $3\times
10^{-6}\Msyr$ (Herrero et al.\ 1995), taken to be constant throughout
the evolution.

The general evolution is reminiscent of that of an 
intermediate-mass neutron-star binary (see Fig.~7 in PRP and the
associated discussion).  After a brief turn-on phase, mass transfer
occurs initially on the thermal timescale of the envelope reaching a
peak mass-transfer rate of $\sim 4\times 10^{-3}\Msyr$.  Once the mass
of the secondary has been reduced to a value comparable to the black
hole, the secondary reestablishes thermal equilibrium and becomes
detached. Indeed because of the continuing wind mass loss the donor
shrinks significantly below its Roche lobe during this phase and the
system widens. The secondary starts to expand again after
it has exhausted all of the hydrogen in the core and fills
its Roche lobe for a second time. In this phase, the mass-transfer
rate reaches a second peak of $\sim 4\times 10^{-4} \Msyr$, where mass
transfer is driven by the evolution of the H-burning shell. The
calculation was terminated at this stage, but the secondary would
ultimately become a $\sim 8\Msun$ helium star, quite possibly becoming
a black hole itself in the final supernova explosion.

The most interesting feature of this calculation is that the system
becomes detached after the initial thermal timescale phase because of
the stellar wind from the secondary (it acts both to widen the orbit
and to shrink the stellar radius). During this phase, mass transfer
continues via the stellar wind. Since the secondary is close to filling its
Roche lobe, such a wind may be focused towards the accreting black
hole, as has been inferred from the tomographic analysis of the mass
flow in Cyg X-1 by Sowers et al.\ (1998). In this particular
calculation, the secondary has a temperature of $\sim 31000\,$K at an
orbital period of 5.6\,d, in excellent agreement with the O9.7
spectral type of HDE 226868. It is worth noting that such a phase will
generally not exist for high-mass neutron-star X-ray binaries since,
because of the more extreme mass ratio in these systems, Roche-lobe
overflow will generally become dynamically unstable leading to the
spiral-in of the neutron star in the envelope of the massive
secondary. If Cyg X-1 is in the phase described above, the model makes
the firm prediction that the secondary should be significantly helium
enriched and its surface composition should show strong evidence for
CNO processing (in this particular model, the surface helium mass
fraction is 0.55, i.e. roughly twice solar, at an orbital period of
5.6\,d). Interestingly, both Herrero et al.\ (1995) and Canalizo et
al.\ (1995) claim to have determined just such abundance anomalies in
the secondary of Cyg X-1, which they argue cannot be accounted for by
uncertainties in the atmosphere modelling.

One caveat is that this model also predicts that the mass of the secondary
should at the present time be comparable to, or lower than, the mass
of the black hole. This does not appear to be consistent with the
analyses of Gies \& Bolton (1986), using rotational velocities and
assuming synchronous rotation, and Gies et al.\ (2002), based on an
H$\alpha$ emission line analysis. Both studies suggest that the
secondary is a factor $\sim 2$\,--\,3 times more massive than the black
hole.

However, irrespective of whether this particular model is applicable
to Cyg X-1, the calculation in Figure~\ref{Fig7} illustrates that it
is generally more likely to observe a high-mass black-hole X-ray
binary in the relatively long-lived wind mass-transfer phase following
the initial thermal timescale phase which only lasts a few $10^4\yr$.
In this example, the wind phase lasts a few $10^5\yr$, but it could last
as long as a few $10^6\yr$ if the secondary were initially less
evolved\footnote{Note that, if the initial mass ratio were more
extreme, the initial mass transfer would become dynamically unstable,
as in the case of massive neutron-star binaries, and the black hole
would spiral into the massive star. In this case, there would be no
subsequent wind mass-transfer phase.}. This means that any secondary
observed in a massive black-hole binary is likely to have already lost a
significant fraction of its mass and that it is generally not
valid to deduce the mass of a secondary based on its spectral type
alone, as is frequently done in the literature.

\subsection{Low-mass black-hole binaries}

\begin{figure*}
\centerline{\psfig{file=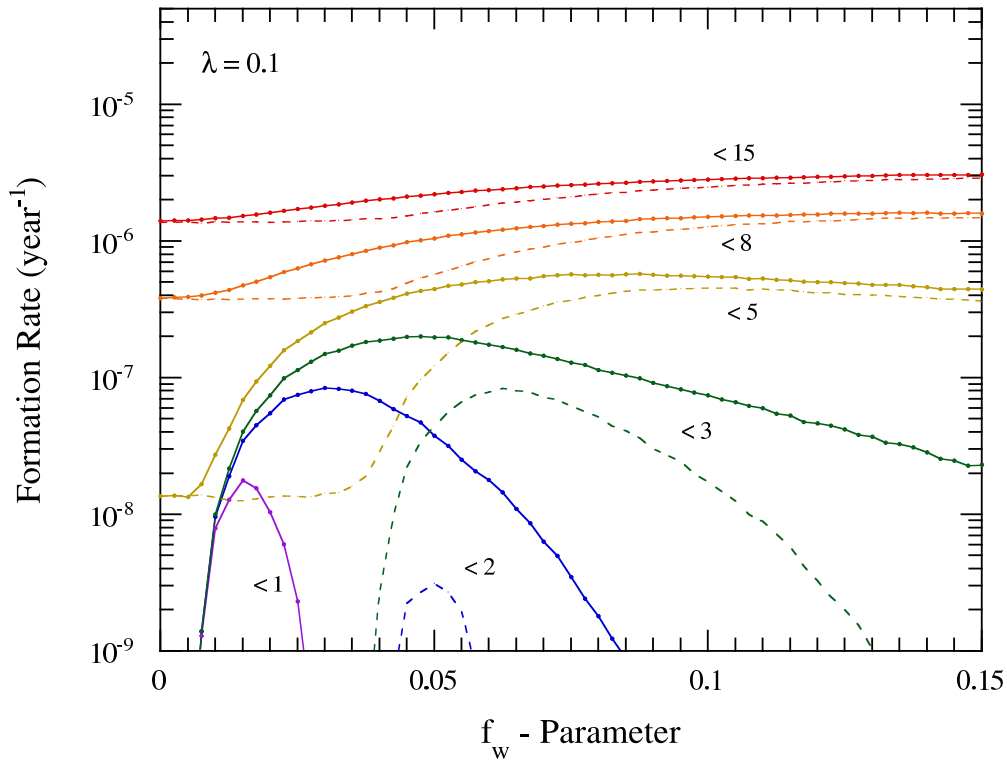,width=10cm,angle=90}}
\caption{
Black-hole binary formation rates as a function of the parameter
$f_w$, the fraction of the primordial primary's stellar wind that is
ejected from the binary with the specific angular momentum of the
primordial {\em secondary}.  The degree of orbital separation (or
contraction) with wind mass loss depends sensitively on this
parameter.  The remaining fraction $(1-f_w)$ is assumed to be lost
with the specific angular momentum of the primary.  The value of the
$\lambda$--parameter has been fixed at an illustrative value of 0.1.
Labels on the different curves refer to the maximum mass of the
secondary star.  Solid curves are {\em without} the inclusion of tidal
interactions between the primary and the orbit. Dashed curves are for
an assumed maximum tidal interaction, i.e. the case where the primary
always rotates synchronously with the orbit.  Note that tidal
interactions typically {\em reduce} the formation rate by tending to
bring systems to Roche-lobe overflow before the primary has lost
sufficient mass to help in the process of unbinding the envelope.
}\label{Fig8}
\end{figure*}

In the binary population synthesis study discussed in \S~2 and the
detailed binary evolution calculations presented in \S~3 hardly any
black-hole binaries were formed with low-mass donors, and all the
systems that were evolved in detail developed into wide, rather than
close binaries.  Thus, 9 of the 17 known black-hole binaries that have
low-mass donors and $P_{\rm orb} \la 1$ day cannot be produced within
the formation scenario we have been evaluating, at least not without
some modification of the input physics we have adopted.  There are two
basic reasons why such low-mass black-hole binaries do not form within
the context of the model presented.  The first has to do with the fact
that the orbital energy in primordial binaries with low-mass
secondaries is generally insufficient to unbind the envelopes of the
massive primaries (see \S~2).  The second reason is that we have
assumed that once a black-hole binary is formed with an intermediate
mass donor, such systems do not experience magnetic braking as a source of
orbital angular momentum loss.  This, in turn, assures that systems
with initially intermediate-mass donors will commence Roche-lobe
overflow (RLOF) only after undergoing a significant amount of nuclear
evolution, subsequent to which the binary evolution necessarily leads
to wider orbits -- including periods considerably longer than a day.
In this section we reevaluate some of our assumptions and relax a
number of constraints to determine whether low-mass, compact,
black-hole binaries can be formed within the standard scenario.

One promising possibility, discussed in \S~2, is that RLOF in the
primordial binary might begin while the primary has a radius between
$R_{\rm HG}$ and $R_{\rm max}$, i.e. after it passes the HG and before it
reaches its maximum radial extent.  During this phase the star is expected
to have lost a significant fraction of its mass in a stellar wind, and this
would greatly reduce the amount of orbital energy required to remove the
residual envelope in a common-envelope phase.  However, as also discussed
in \S~2, the mass loss tends to make the orbit expand even faster than the
primary can expand, and hence RLOF is not likely to occur.  In this case, a
close binary will not be formed, if it remains bound at all.  The
calculations that led to this conclusion involved the assumptions that (1)
the specific angular momentum carried away by the wind of the primary has
the same value as that of the primary itself, $j_p$, and (2) no significant
synchronizing tidal torques act between the expanding primary and the
orbit.

If we suppose that a fraction, $f_w$, of the primary's stellar wind is
deflected by the secondary and leaves the binary with the larger specific
angular momentum of a low-mass secondary, $j_s$, then the orbital expansion
will be diminished, and may even be reversed.  This, in turn, could enable
RLOF to occur when the primary has lost a significant fraction of its
envelope.  We have carried out a series of BPS calculations where the
specific angular momentum carried away by the wind of the primary is set
equal to
\begin{equation}
j_w = (1-f_w) j_p + f_w j_s  .
\end{equation}
The results for the black-hole binary formation rates vs. the fraction
$f_w$ are shown as solid curves in Figure~\ref{Fig8}.  These were
calculated for an assumed, illustrative, fixed value of $\lambda =
0.1$.  As in Figure~\ref{Fig3} the different curves are for various
limits on the mass of the donor star in successfully formed black-hole
binaries.  Note that for $f_w = 0$, the results agree with those in
Figure~\ref{Fig3}, and show that black-hole binaries with donors $\la
6 M_\odot$ hardly form with such a small value of $\lambda$.  However,
as the value of $f_w$ is increased by a small amount, e.g. to $\sim
0.05$, the formation rates of systems with lower-mass donors grow
significantly.  For still larger values of $f_w$, the formation rates
drop back down again.  This can be understood as follows.  For small
values of $f_w$ the orbital expansion can be substantially limited,
and the primary has a chance to shed a significant part of its
envelope before the common-envelope phase.  For larger values of $f_w$
the orbit actually {\em shrinks} sufficiently rapidly that RLOF may
commence {\em before} much envelope mass can be lost in a wind.  The
main problem with this hypothesis (invoking extra angular-momentum
loss) is that for lower-mass secondaries the fraction of the wind
deflected by the secondary goes as the square of the mass ratio, and
such fractions are typically much smaller than the values of $f_w$ for
which this effect is important (however, see, Hachisu, Kato \& Nomoto
1999).

We have carried out several other related exploratory tests in which the
wind loss rate from the primary was taken to be lower than that given by
the Nieuwenhuijzen \& de Jager (1990) prescription by factors of 2 and 5.
This has a similar effect on inhibiting the orbital widening as discussed
above when the specific angular momentum of the wind was enhanced.
However, for neither diminution factor of the wind loss rate were a
significant number of low-mass black-hole binaries formed for any value of
$\lambda$.

\begin{figure*}
\centerline{\psfig{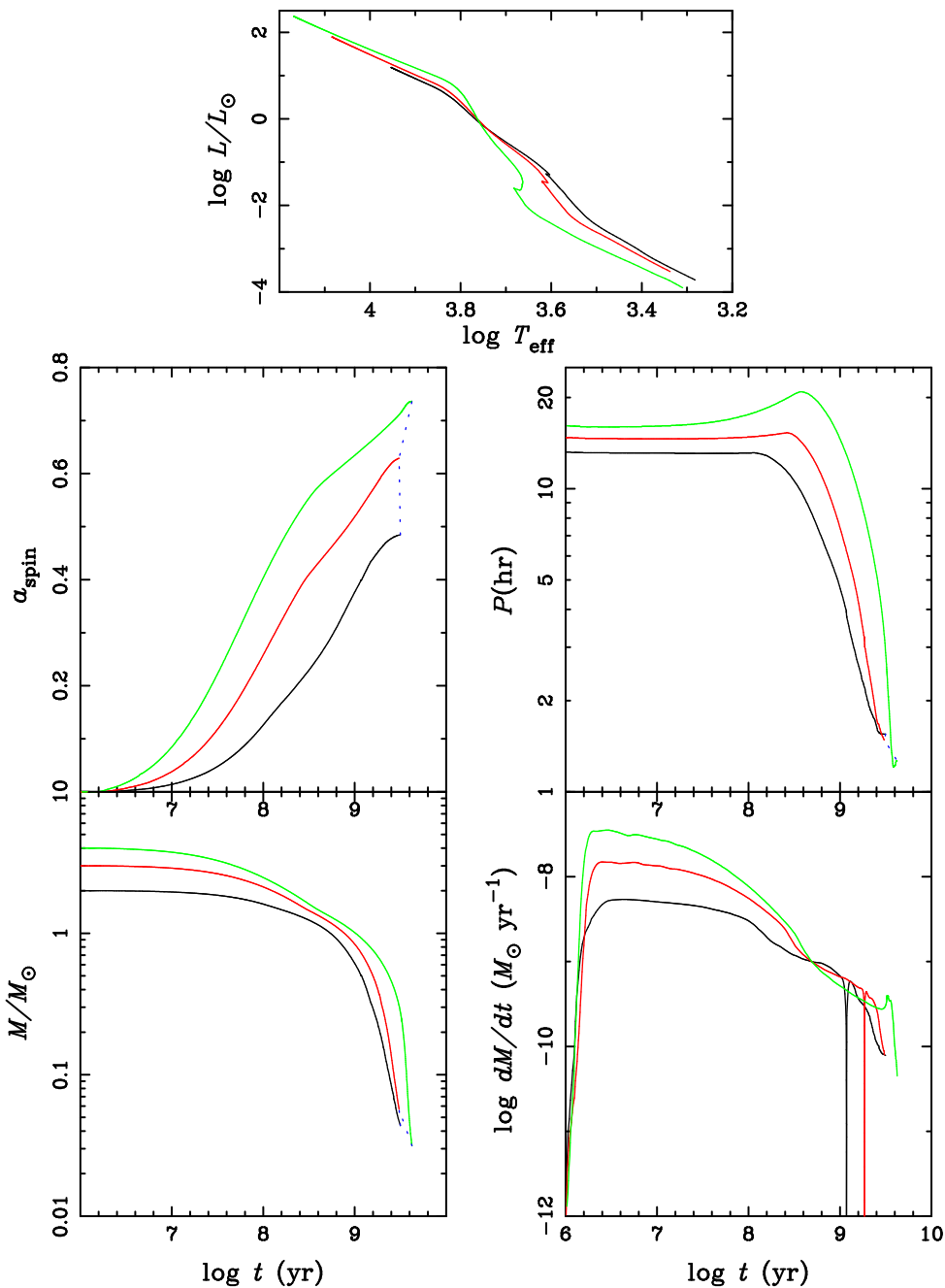}}
\caption{H-R diagram (top panel) and key binary parameters
as a function of time since the beginning of mass transfer
(with arbitrary offset) for binary sequences with initial masses
of 2\Msun\ (solid curves), 3 \Msun\ (dashed curves) and 4 \Msun\
(dot-dashed curves) where magnetic braking has been included even for
secondaries with radiative envelopes.
{\it Middle left:} Spin parameter. {\it Middle right:} Orbital period.
{\it Bottom left:} Mass of the secondary. {\it Bottom right:}
Mass-loss rate of the secondary.}\label{Fig9}
\end{figure*}

Finally, in this regard we tested the effects of introducing a strong
tidal coupling between the orbit and the rotation of the primary as it
evolves to become a giant.  We considered the extreme case where the
primary remains in corotation with the orbit as it expands through the
giant phase.  This is equivalent to assuming that the tidal
synchronization timescale is always short compared to the nuclear
evolution timescale of the primary -- which may not be unreasonable for
these largely convective stars (see e.g. Lecar, Wheeler \& McKee
1976; Zahn 1977; Verbunt \& Phinney 1995).  The tidal
interactions have the effect of slowing the orbital expansion or
causing orbital contraction as the primary's moment of inertia grows,
and in the case of low-mass secondaries the Darwin instability is even
likely to set in.  In the latter case, we simply assume that RLOF
commences at that point.  The results of our BPS calculations for
strong tidal interactions are shown in Figure~\ref{Fig8} as a set of
dashed curves (same mass color coding as for the no tidal interaction
case).  These represent the formation rates for black-hole binaries
with different secondary masses -- again as a function of the $f_w$
parameter discussed above.  For systems with higher mass secondaries,
the assumed tidal interaction does not significantly change the
formation rates.  However, for the lower mass secondaries, it is clear
that tidal interactions serve only to {\em decrease} the formation
rates, rather than enhance them.  This is due to the fact that for small
values of $q = M_{\rm s}/M_{\rm p}$, the tidal influence is quite large, leading
to either rapid orbital shrinkage or a runaway Darwin instability.
This forces an earlier commencement of RLOF and leaves much of the
primary's envelope to be ejected in the common envelope.

The net conclusion of these studies is that it is still quite
problematic to form low-mass black-hole binaries directly via a
common-envelope scenario.  If, on the other hand, systems with
initially intermediate-mass donors could evolve to short orbital
periods with low-mass donors, then initially low-mass donors might not
be required at all. Indeed, as was shown in PRP, intermediate-mass
X-ray binaries containing neutron stars rather than black holes can be
almost indistinguishable from low-mass systems after an initial
thermal timescale mass-transfer phase. Even systems with secondaries
as high as 3.5\Msun\ can evolve towards short orbital periods if the
secondary is initially relatively unevolved. The reason is that, for
intermediate-mass neutron-star binaries, mass transfer initially
occurs from the more massive to the less massive component of the
system which leads to a decrease in the orbital period. When the
secondaries develop convective envelopes, magnetic braking becomes the
dominant mechanism to drive mass transfer, causing the orbit to shrink
further. In the intermediate-mass black-hole binaries, the initial
mass ratio is reversed and the systems tend to widen from the
beginning. This situation would, however, be different if there were
an additional angular-momentum loss mechanism operating in these
systems.

To test this possibility, we decided to relax our condition for when
orbital angular momentum losses via magnetic braking of the donor star
are effective.  In particular, we investigated how black-hole binaries
with intermediate-mass donor stars would evolve if magnetic braking
were operable, independent of the presence or absence of a convective
envelope.  In Figure~\ref{Fig9} we show detailed binary evolution
results for systems with donor stars of initial mass 2, 3 and 4\Msun\
with the inclusion of continuous magnetic braking. Note that within
some $\sim 10^9$ yr of the commencement of mass transfer, the mass of
all the donors has been reduced below $\sim 1\Msun$ and that the
orbital periods have decreased from their initial values of $\sim 1/2$
day through a range of several hours. The calculations were terminated
when the systems reached their period minimum, which varied from
76\,min for the initially 4\Msun\ donor to 93\,min for the initially
2\Msun\ donor (the final composition in the former case is
significantly hydrogen-depleted and has a hydrogen mass fraction of
0.34, while in the latter case the secondary is only moderately hydrogen
depleted).

We have no direct way of evaluating whether conventional ideas about
magnetic braking can be overturned in this way or whether there might
be some other angular-momentum loss mechanism in order to form the
observed low-mass black hole binaries. However, it seems to us to be a
more attractive alternative than invoking common-envelope scenarios
that implicitly violate conservation of energy.

\subsection{Discussion and Conclusions}

In this paper we systematically explored the formation and evolution
of black-hole binaries. Our binary calculations have shown that mass
transfer is stable for a wide range of donor masses (up to about $\sim
20\Msun$ for an initial black-hole mass of 10\Msun).  After sufficient
mass has been lost from the donor, an initially intermediate-mass
donor can mimic a low-mass one and a high-mass donor can mimic an
intermediate-mass one. We have shown that black holes can gain
substantial mass from the companion even if accretion is
Eddington-limited and that the black hole can be spun up
significantly. This demonstrates that the present black-hole mass is
not necessarily representative of the initial black-hole mass after
the supernova in which it formed, an important fact to be taken into
account when studying the implications of observed black-hole masses
for single-star evolution.

Our models are directly applicable to many observed black-hole
binaries, where we particularly concentrated on GRS 1915+105 and Cyg
X-1. Our calculations show that the initial mass of the mass donor in
GRS 1915+105 may have been as high as $\sim 6\Msun$ and the black hole
may have accreted up to $\sim 4\Msun$ from its companion, being spun
up in the process. The composition of the donor, in particular the
helium abundance and the amount of CNO processing it underwent,
provides a potentially powerful test of the initial mass of the donor
star. V404 Cyg, a massive black-hole binary with an orbital period
of 6.5\,d and some of the best determined system parameters (Shahbaz
et al.\ 1994, 1996), is also very well reproduced by our evolutionary
sequences.

Our calculations may also help us to understand the nature of
ultraluminous X-ray sources (ULXs) in external galaxies, in particular
those found in regions of active star formation. In our massive
sequences, the systems reach {\em potential\/} X-ray luminosities as
high as $\sim 10^{41}\ergss$, comparable to the most luminous ULXs
observed, when the donor stars become giants and the evolution is
driven by the nuclear evolution of the hydrogen-burning shell.

We have also performed detailed binary population synthesis
calculations which show that intermediate- and high-mass black-hole
binaries can form at reasonable rates with plausible assumptions,
while (initially) low-mass black-hole binaries apparently cannot,
unless one is prepared to accept violation of energy conservation. We
have explored various possible solutions to reconcile this conclusion
with the large number of low-mass, short-period black-hole binaries
observed. These include: (1) The possibility that there are serious
flaws in the stellar models of massive, evolved stars, or in the
assumptions concerning the ejection of the common-envelope should be
considered. (2) The assumptions concerning the angular-momentum loss
in the stellar wind from the primary have to be changed drastically.
(3) Low-mass, short-period systems are in fact descendants of
intermediate-mass systems. This requires an additional
angular-momentum loss mechanism in systems with intermediate-mass
radiative stars (e.g. continuous magnetic braking).  (4) Low-mass
systems form through one of the alternative evolutionary formation
channels suggested, e.g. in a triple scenario or out of the collapsed
envelope of a massive star.

Finally, we conclude that before some of these fundamental issues have been
resolved, the predictions of binary population synthesis studies of
black-hole binaries have to be taken with considerable caution.

\section*{Acknowledgements}

We thank E. Pfahl, R. Remillard, P.A. Charles, I.F. Mirabel
for extremely helpful discussions and the referee, V. Kalogera, for 
very useful comments.
This work was in part supported by a Royal Society UK-China Joint Project
Grant (Ph.P and Z.H.), the Chinese National Science Foundation under
Grant No.\ 19925312, 10073009 and NKBRSF No. 19990754 (Z.H.) and by the
National Aeronautics and Space Administration under ATP grant NAG5-8368.


\begin{thebibliography}{}

\bibitem[]{} Abt H.A., Levy S.G., 1978, ApJS, 36, 241
\bibitem[]{} Angelini L., Loewenstein M., Mushotzky R.F., 2001, ApJ, 557, L35
\bibitem[]{} Alexander D.R., Ferguson J.W., 1994, ApJ, 437, 879
\bibitem[]{} Arzoumanian Z., Chernoff D.F., Cordes J.M., 2002, ApJ,
568, 289 
\bibitem[]{} Bardeen J.M., 1970, Nat, 226, 64
\bibitem[]{} Begelman M.C., 2002, ApJ, submitted (astro-ph/0203030)
\bibitem[]{} Belczynski K., Bulik T., 2002, submitted (astro-ph/0205248)
\bibitem[]{} Brandt W.N., Podsiadlowski Ph., 1995, MNRAS, 274, 461 
\bibitem[]{} Brandt W.N., Podsiadlowski Ph., Sigurdsson S., 1995, MNRAS,
277, L35
\bibitem[]{} Brown G.E., Heger A., Langer N., Lee C.-H.,
Wellstein S., Bethe H.A., 2001, New Astr., 6, 457
\bibitem[]{} Brown G.E., Lee C.-H., Bethe H.A., 1999, New Astron., 4, 313
\bibitem[]{Brown} Brown G.E., Lee C.-H., Wijers R.A.M.J., Bethe H.A.,
2000, Physics Reports, 333, 471
\bibitem[]{} Canalizo G., Koenigsberger G., Pe\~na D., Ruiz E., 1995,
Rev.\ Mexicana Astron.\ Astrofis., 31, 63
\bibitem[]{} Cannizzo, J., Wheeler, J.C., 1984, ApJS, 55, 367.
\bibitem[]{Cappellaro} Cappellaro E., Evans R., Turatto M. 1999, A\&A,
351, 459
\bibitem[]{} Castro-Tirado A.J., Brandt S., Lund N., 1992, IAU Circ., 5590
\bibitem[]{} Chevalier R.A., 1993, ApJ, 411, L33
\bibitem[]{Chiang} Chiang E., Rappaport S., 1996, ApJ, 469, 255
\bibitem[]{} Colbert E.J.M, Mushotzky R.F., 1999, ApJ, 519, 89
\bibitem[]{} Colbert E.J.M., Ptak A.F., 2002, submitted (astro-ph/0204002)
\bibitem[]{} de Kool M., 1990, ApJ, 358, 189
\bibitem[]{} de Kool M., van den Heuvel E.P.J., Pylyser E., 1987, A\&A, 183, 47
\bibitem[]{} Dewi J., Tauris T., 2000, A\&A, 360, 1043
\bibitem[]{} Dewi J., Tauris T., 2001, in Podsiadlowski Ph., Rappaport S.,
King A.R., D'Antona F., Burderi L., eds, Evolution of Binary and
Multiple Stars Systems, ASP Conf. Ser., Vol 229, 255
\bibitem[]{Duquennoy} Duquennoy A., Mayor M., 1991, A\&A, 248, 485
\bibitem[]{} Eggleton P.P., Verbunt F., 1986, MNRAS, 220, 13
\bibitem[]{} Ergma E., Fedorova A., 1998, A\&A, 338, 69
\bibitem[]{} Ergma E., van den Heuvel E.P.J., 1998, A\&A, 331, L29
\bibitem[]{} Fabbiano G., 1989, ARA\&A, 27, 87
\bibitem[]{} Fryer C.L., Burrows A., Benz W., 1997, ApJ, 496, 333
\bibitem[]{} Fryer C.L., Heger A., 2000, ApJ, 541, 1033 
\bibitem[]{} Fryer C.L., Kalogera V., 2001, ApJ, 554, 548
\bibitem[]{Garmany} Garmany C.D., Conti P.S., Massey P., 1980, ApJ,
242, 1063
\bibitem[]{} Gies D.R. et al., 2002, ApJ, submitted (astro-ph/0206253)
\bibitem[]{} Gies D.R., Bolton C.T., 1982, ApJ, 260, 240
\bibitem[]{} Gies D.R., Bolton C.T., 1986, ApJ, 304, 371
\bibitem[]{} Greiner J., Cuby J.G., McCaughrean M.J., 2001, Nat, 414, 522
\bibitem[]{} Greiner J., Morgan E.H., Remillard R.A., 1996, ApJ, 473, L107
\bibitem[]{} Greiner J., Morgan E.H., Remillard R.A., 1997,
in Ogley R.N., Bell-Burnell, J., eds, Galactic Sources with
Relativistic Jets, New Astron.\ Rev., 597
\bibitem[]{} Hachisu I., Kato M., Nomoto K., 1999, ApJ, 522, 487
\bibitem[]{} Han Z., Podsiadlowski Ph., Eggleton P.P., 1994, MNRAS, 270, 121
\bibitem[]{} Han Z., Podsiadlowski Ph., Maxted P.F.L., Marsh T.R., Ivanova N.,
2002a, MNRAS, accepted (astro-ph/0206130)
\bibitem[]{} Han Z., Podsiadlowski Ph., Maxted P.F.L., Marsh T.R.,
2002b, MNRAS, submitted
\bibitem[]{Hansen} Hansen B.M.S., Phinney E.S., 1997, MNRAS, 291, 569
\bibitem[]{} Herrero A., Kudritzki R.P., Gabler R., Vilchez J.M., Gabler A.,
1995, A\&A, 297, 556
\bibitem[]{} Hjellming M.S. ,Webbink R.F., 1987, ApJ, 318, 794
\bibitem[]{} Houck J.C, Chevalier R.A., 1991, ApJ, 376, 234
\bibitem[]{Hurley} Hurley J.R., Pols O.R., Tout C.A., 2000, MNRAS,
315, 543 (HPT)
\bibitem[]{} Janka H.-T., M\"uller E., 1994, A\&A, 290, 496 
\bibitem[]{} Jeltema J.E., Canizares C.R., Buote, D.A., Garmire G.P.,
2002, ApJ, submitted
\bibitem[]{} Kalogera V., 1999, ApJ, 521, 723
\bibitem[]{} Kalogera V., Webbink R.F., 1996, ApJ, 458, 301
\bibitem[]{} King A.R., 2002, MNRAS, submitted (astro-ph/0206117)
\bibitem[]{} King A.R., Davies M.B., Ward M.J., Fabbiano G.,
Elvis M., 2001, ApJ, 552, L109
\bibitem[]{} King A.R., Kolb U., 1999, MNRAS, 305, 654
\bibitem[]{} King A.R., Kolb U., Szuskiewicz E., 1997, ApJ, 488, 89
\bibitem[]{} Kippenhahn R., Weigert A., 1990, Stellar Structure and
Evolution (Springer, Berlin)
\bibitem[]{} Kippenhahn R., Weigert A., Hofmeister E., 1967 in Alder B.,
Fernbach S., Rothenberg M., eds, Methods in Computational Physics,
Vol.~7, p.\ 129
\bibitem[]{} Kundu A., Maccarone T.J., Zepf S.E., 2002, ApJ, submitted
(astro-ph/0206221)
\bibitem[]{} Lai D. 2000, in Blaschke D., Glendenning N.K., 
\& Sedrakian, A., eds, Physics of Neutron Star Interiors (Springer) 
(astro-ph/0012049)  
\bibitem[]{} Langer N., Maeder A., 1995, A\&A, 295, 685
\bibitem[]{} Lasota J.-P., 2002, New Astronomy Reviews, 45, 449
\bibitem[]{} Lee C.-H., Brown G.E., Wijers R.A.M.J., 2002, submitted
(astro-ph/0109538) (LBW)
\bibitem[]{} Lecar M., Wheeler J. C., McKee C. F. 1976, ApJ, 205, 556
\bibitem[]{Levine91} Levine A., Rappaport S., Putney A., Corbet R.,
Nagase F., 1991, ApJ, 381, 101
\bibitem[]{Levine93} Levine A., Rappaport S., Deeter J., Boynton P.,
Nagase F., 1993, ApJ, 410, 328
\bibitem[]{} Lira P., Johnson R., Lawrence A., 2002, MNRAS, submitted
(astro-ph/0206123)
\bibitem[]{} Lyne A.G., Lorimer D.R., 1994, Nature, 369, 127 
\bibitem[]{} Makishima K., et al., 2000, ApJ, 535, 632
\bibitem[]{} Maeder A., 1992, A\&A, 264, 105
\bibitem[]{Miller} Miller G.E., Scalo J.M., 1979, ApJS, 41, 513
\bibitem[]{} Mirabel I.F., Rodr\'\i guez L.F., 1994, Nat, 371, 46
\bibitem[]{} Mirabel I.F., Rodr\'\i guez L.F., 1999, ARA\&A, 37, 409
\bibitem[]{} Nelemans G., van den Heuvel E.P.J., 2001, A\&A, 376, 950
\bibitem[]{} Nelemans G., Tauris T.M., van den Heuvel E.P.J., 1999, A\&A, 352,
87
\bibitem[]{Nieuwenhuijzen} Nieuwenhuijzen, H., de Jager, C., 1990, A\&A,
231, 134
\bibitem[]{} Nugis T., Lamers H.J.G.L.M., 2000, A\&A, 360, 227
\bibitem[]{Orosz} Orosz J.A., et al., 2002, ApJ, accepted (astro-ph/0112101)
\bibitem[]{} Paczy\'nski B., 1976, in Eggleton P.P., Mitton S., Whelan J.,
eds, Structure and Evolution of Close Binaries (Kluwer, Dordrecht), 75
\bibitem[]{Pakull} Pakull M., Mirioni L., 2002, in  Jansen F. et al., eds,
New Visions of the X-ray Universe in the XMM-Newton and Chandra
Era, 26-30 (astro-ph/0202488)
\bibitem[]{Pfahl} Pfahl E., Rappaport S., Podsiadlowski Ph., Spruit H.,
2002, ApJ, in press (astro-ph/0109521)
\bibitem[]{Plavec} Plavec M., 1968, Adv.\ A\&A, 6, 201
\bibitem[]{} Podsiadlowski Ph., 2001, in Podsiadlowski Ph.,
Rappaport S., King A.R., D'Antona F., Burderi L., eds, Evolution
of Binary and Multiple Star Systems, ASP Conf.\ Ser., Vol.\ 229, P. 239
\bibitem[]{} Podsiadlowski Ph., Cannon R.C., Rees M.J., 1995, MNRAS, 274, 485
\bibitem[]{} Podsiadlowski Ph., Nomoto K., Maeda K.,  Nakamura T.,  Mazzali
P., Schmidt B., 2002a, ApJ, 567, 491
\bibitem[]{} Podsiadlowski Ph., Rappaport S., 2000, ApJ, 529, 946
\bibitem[]{} Podsiadlowski Ph., Rappaport S., Pfahl E., 2002b, ApJ, 565, 1107
(PRP)
\bibitem[]{} Pols O.R., Dewi J.D.M., 2002  (astro-ph/0203308)
\bibitem[]{} Pols O.R., Tout C.A., Schr\"oder K.-P., Eggleton P.P., Manners,
J., 1997, MNRAS, 289, 869
\bibitem[]{} Portegies Zwart S.F., Verbunt F., Ergma E., 1997, A\&A, 321, 207
\bibitem[]{} Prestwich A.H., Irwin J.A., Kilgard R.E., Krauss M.I., Zezas A.,
Primini F., Kaaret P., ApJ, submitted (astro-ph/0206127)
\bibitem[]{} Rappaport S., Verbunt F., Joss, P.C. 1983, ApJ, 275, 713
\bibitem[]{} Ritter H., 1996, in Evolutionary Processes in Binary
Stars, Wijers R.A.M.J., Davies M.B., Tout C.A., eds,
(Kluwer, Dordrecht), p. 223
\bibitem[]{} Roberts T.P., Goad M.R., Ward M.J., Warwick R.S., Lira P., 2002,
in Jansen F. et al., eds, New Visions of the X-ray Universe in the
XMM-Newton and Chandra Era, in press (astro-ph/0202017)
\bibitem[]{} Roberts T.P., Warwick R.S., 2000, MNRAS, 315, 98
\bibitem[]{} Rogers F.J., Iglesias C.A., 1992, ApJS, 79, 507
\bibitem[]{} Romani R.W., 1992, ApJ, 399, 621
\bibitem[]{} Romani R.W., 1998, A\&A, 333, 583
\bibitem[]{Salpeter} Salpeter E.E., 1955, ApJ, 121, 161
\bibitem[]{} Taam R.E., Sandquist E.L., 2000, ARA\&A, 38, 113
\bibitem[]{} Sazonov S.Y., Sunyaev R.A., Lapshov I.Y., Lund N., Brandt S.,
Castro-Tirado A., 1994, Pisma V. Astron.\ Zhurnal V., 20, No. 12, 901
\bibitem[]{} Schr\"oder K.-P., Pols O.R., Eggleton P.P., 1997, MNRAS, 285, 696
\bibitem[]{} Shahbaz T., Ringwald F.A., Bunn J.C., Naylor T., Charles P.A.,
Casares J., 1994, MNRAS, 271, L10
\bibitem[]{} Shahbaz T., Bandyopadhyay R., Charles P.A., Naylor T., 1996,
MNRAS, 282, 977
\bibitem[]{} Sowers J.W., Gies D.R., Bagnuolo Jr. G., Shafter A.W., Wiemker R.,
Wiggs M.S., 1998, ApJ, 506, 424
\bibitem[]{Spruit} Spruit H.C., Phinney E.S., 1998, Nat, 393, 139
\bibitem[]{} Strai\v{z}ys V., Kuriliene G., 1981, Ap\&SS, 80, 353
(Dordrecht: Kluwer)
\bibitem[]{} Tauris T.M., Dewi J.D.M., 2001, A\&A, 369, 170
\bibitem[]{} Terashima Y., Wilson A.S., 2002, in New Visions of the X-ray
Universe in the XMM-Newton and Chandra Era, Jansen F. et al., eds, in press
(astro-ph/0204321)
\bibitem[]{Thorne} Thorne K.S., 1974, ApJ, 191, 507
\bibitem[]{} van den Heuvel E.P.J., 2001, in Podsiadlowski Ph., 
Rappaport S., King A.R., D'Antona F., Burderi L., eds, 
Evolution of Binary and Multiple Star Systems, ASP Conf.\ Ser., Vol.\ 229, 525
\bibitem[]{} van den Heuvel E.P.J., Portegies Zwart S.F., Bhattacharya D., 
Kaper L., 2000, A\&A, 364, 563 
\bibitem[]{} van Paradijs J., 1996, ApJ, 464, L139
\bibitem[]{} Verbunt F., Phinney E.S., 1995, A\&A, 296, 709
\bibitem[]{} Verbunt F., van den Heuvel, E.P.J., 1995, in Lewin W.H.G., 
van Paradijs J., van den Heuvel E.P.J., eds, X-ray Binaries 
(Cambridge University Press: Cambridge), p. 457
\bibitem[]{} Verbunt F., Zwaan, C., 1981, A\&A, 100, L7
\bibitem[]{} Vilhu O., 2002, submitted (astro-ph/0204146)
\bibitem[]{} Vrtilek, S.D., Raymond, J.C., Garcia, M.R., Verbunt, F., 
Hasinger, G., Kurster, M., 1990, A\&A, 235, 162
\bibitem[]{} Walborn N.R., 1973, ApJ, 179, L123
\bibitem[]{} Wellstein S., Langer N., 1999, A\&A, 350, 148
\bibitem[]{White} White N.E., van Paradijs J., 1996, ApJ, 473, L25.
\bibitem[]{} White III R.E., Sarazin C.L., Kulkarni S.R., 2002, ApJ, submitted
(astro-ph/0204172)
\bibitem[]{} Wijers R.A.M.J., 1996 in Evolutionary Processes in Binary
Stars, Wijers R.A.M.J., Davies M.B., Tout C.A., eds,
(Kluwer, Dordrecht), p. 327
\bibitem[]{} Woosley S.E., Weaver T.A., 1995, ApJS, 101, 181
\bibitem[]{} Woosley S.E., Langer N., Weaver T.A., 1995, ApJ, 448, 315
\bibitem[]{} Zahn, J.-P. 1977, A\&A, 57, 383
\bibitem[]{} Zezas A., Fabbiano G., Rots A.H., Murray S.S., 2002,
ApJ, submitted (astro-ph/0203175)
\bibitem[]{} Zhang S.N., Cui W., Chen W., 1997, ApJ, 482, L155
\label{lastpage}
\end{thebibliography}
\end{document}